\documentclass[%
twocolumn,
amsmath,amssymb,
prb,
]{revtex4-1}

\usepackage{graphicx,enumitem}% Include figure files
\usepackage{dcolumn}% Align table columns on decimal point
\usepackage{bm}% bold math
\usepackage{placeins}

\usepackage{textcomp}

\begin{document}

\preprint{APS/123-QED}

\title{First principles calculations of band offsets at heterovalent $\varepsilon$-Ge/In$_x$Al$_{1-x}$As interfaces}
% :\\with Forced Linebreak}% Force line breaks with \\
\newcommand{\etsf}{European Theoretical Spectroscopy Facilities (ETSF)}
\newcommand{\qub}{School of Mathematics and Physics, Queen's University Belfast, Belfast BT7 1NN, Northern Ireland, UK}
\author{G. Greene-Diniz}
\email{g.greene-diniz@qub.ac.uk}
\affiliation{\qub}
\author{M. Gr\"{u}ning}
\affiliation{\qub}
\affiliation{\etsf}


%\affiliation{
%\affiliation{
%\affiliation{%


\date{\today}% It is always \today, today,

\begin{abstract}
First-principles electronic structure calculations are carried out to investigate the band alignments of tensile strained (001) Ge interfaced with (001) In$_{x}$Al$_{1-x}$As. The sensitivities of band offsets to interfacial structure, interfacial stoichiometry, and substrate stoichiometry, are investigated. Large qualitative variations of the valence and conduction band offsets are observed, including changes of the band offset type, indicating the importance of local structural variations of the interface for band offsets in real samples. Our results explain recent measurements of band offsets derived from XPS core level spectra in terms of As atoms penetrating through the first few monolayers of the Ge film. Analogous studies are carried out for the diffusion of other species across the interface, and in general, the band offsets vary approximately linearly with diffusion depth relative to the values for pristine ``sharp'' interfaces, where the sign of the linear variation depends on the diffusing species. This large sensitivity of the band alignments to interface details indicates potential routes to chemically control the band offset of this group-IV/III-V interface by tuning the stoichiometry of the substrate surface that the thin-film is grown on.
%\begin{description}
%\end{description}
\end{abstract}

\maketitle

%\tableofcontents

\section{\label{sec:level1}Introduction\\}

The interface of tensile strained germanium ($\varepsilon$-Ge) grown on III-V substrates is currently being considered as the working tunnel-barrier in the channel of future high-performance and low-power consumption tunnel field-effect transistors (TFETs).\cite{Hudait2014,Clavel2015,Nguyen2015} These devices take advantage of band-to-band tunneling of charge carriers between the source and drain, and as a result, can overcome the limit for the subthreshold slope of thermionic devices,\cite{Ionescu11} thereby simultaneously improving the transistor switching speed (performance) and I$_\text{ON}$/I$_\text{OFF}$ current ratio (power efficiency). Concurrently, there is a large research effort dedicated to the integration of optical interconnects on a CMOS compatible platform,\cite{Miller09,Assefa10} allowing for highly efficient ultrafast inter- and intra-chip data communication. The latter requires efficient on-chip light sources, and $\varepsilon$-Ge grown on III-V substrates\cite{Pavarelli13} is being investigated for this purpose due to the tensile strain induced direct band gap of Ge.

The operation of transistor devices depends crucially on the junctions at the border between device materials, and this dependence only becomes stronger as device dimensions continue to shrink.\cite{delAlamo11,Ferain11} From the perspective of optical devices, where electron-hole recombination is required in the active region for light emission, material interfaces also play a dominant role in the device operation\cite{Pavarelli13} by determining the barrier height for electron and hole confinement. Motivated by the technological importance, significant progress has been made in recent decades towards the understanding of solid-state interfaces and the resulting line up of energy bands between materials forming the interface.\cite{Peressi98,Margaritondo12,Agostini13,Brillson16} 

In terms of \textit{ab-initio} calculations of band alignments, the lattice-matched isovalent interfaces\cite{Vandewalle87,Vandenberg88,Christensen88,Peressi90,Hybertsen90,Hybertsen91} represent the simplest and most well studied category. Density functional theory (DFT) is typically employed, and idealized, atomically abrupt interfacial structures\cite{Harrison78,Peressi98} are often used. These theoretical works have shown that band alignments for these interfaces are predominantly derived from bulk properties of the adjoining materials.\cite{Vandewalle87,Christensen88} Hence, for isovalent interfaces, the interfacial structure does not have a significant effect. 

Band offsets (BOs) across pseudomorphic heterostructures exhibiting heterovalent bonding across the interface have also been studied, both experimentally\cite{Waldrop79,Kraut80,Biasol92,Dahmen93,Volodin14,Pavarelli13} and computationally.\cite{Martin80,Kunc81,Peressi91,Biasol92,Franciosi93,Peressi98,Pavarelli13} As for the isovalent interfaces, a large portion of the computational (atomistic modeling) studies also involve ideal, abrupt interfaces, although some focus has been given to atomic intermixing/diffusion across the interface.\cite{Peressi98} Unlike isovalent junctions, the interfacial structure can have a significant effect on the band offsets of heterovalent interfaces, to the point of inducing \textit{qualitative} modifications to the offsets, e.g. type-I to type-II or vice versa, in interfaces such as $\varepsilon$-Ge/In$_{0.3}$Ga$_{0.7}$As(001),\cite{Pavarelli13} and also for Ge/In$_{x}$Al$_{1-x}$As(001) in certain cases (see results section). 

In this paper, the sensitivity of BOs to interface structure in the lattice (mis)matched heterovalent ($\varepsilon$-)Ge/In$_{x}$Al$_{1-x}$As(001) interface is explained by a linear response electrostatic effect\cite{Resta89,Peressi98} which occurs as a result of changes in the position of polarized bonds (IV-III, or IV-V) relative to the abrupt interface. Local changes in the electrostatic potential step across the junction result from the local variations in valence charge density and the latter are in turn induced by variations in the stoichiometry of the interfacial region. Hence, this work extends previous theories of BO-interface structure relations\cite{Resta89,Peressi98} to the technologically important interface ($\varepsilon$-)Ge/In$_{x}$Al$_{1-x}$As(001). By explaining the qualitative changes in the band alignments that can be achieved for the same material interface by only changing the interface structure, this work also contributes to the understanding of how devices can be tailored by the interface.

Lattice mismatching across the interface can also affect the band alignment. When a thin-film is grown pseudomorphically on a substrate with a different lattice constant the thin-film exhibits elastic strain so that it can match the lattice constant of the substrate, below a critical thickness such that stress is not large enough to cause plastic relaxation via e.g. dislocation formation.\cite{ChasonGuduru16} Epitaxial strain can be used to induce a direct bandgap in Ge, useful for silicon-compatible photonics.\cite{LiangBowers10} For strained, heterovalent interfaces such as ($\varepsilon$-)Ge/In$_x$Al$_{1-x}$As the dependence of the band alignment on tensile strain $\varepsilon$, which is varied by the substrate stoichiometry $x$, can be significant due to the reordering of conduction band valleys. Here, we study the band alignment over a range of cation stoichiometry $x$ and show that when combined with modifications of the interface structure (modifications which represent diffusion of group-III atoms into the Ge layer), transitions between type-I and type-II band alignments can be achieved in this interface. 

In this work, the variations of valence (VBO) and conduction (CBO) band offsets between Ge and In$_{x}$Al$_{1-x}$As, with respect to interfacial configuration, are investigated using first-principles atomistic simulations---which are detailed in the next section. In Sec.~\ref{results}, we consider a range of systematic structural modifications of the interface. Specifically, we consider (a) the Ge, As, and group-III stoichiometric balance of the mixed interfacial region for fixed substrate stoichiometries, (b) group-III composition of the In$_{x}$Al$_{1-x}$As substrate (for $x$ = 0.0 to 0.25) for fixed interfacial stoichiometries (Sec.~\ref{sec-abrupt-int}), (c) interdiffusion of species across the junction (Sec.~\ref{sec-intdiff}). For interdiffusion, we investigate the relative stability of diffused atoms in either material (Sec.~\ref{sec-Eform}). Finally, in Sec.~\ref{sec-analysis}, we rationalize the results by simple arguments and models based on the linear response electrostatic effect. Based on the results of the simulations and on the linear response analysis, we conclude  (Sec.~\ref{conclusions}) that our simulations provide a picture consistent with existing experimental results. We predict that both type-I and type-II band offsets should be observable for this interface depending on the details of the interface structure.

\section{Computational Methods}
Optimized geometries of bulk and interface models are calculated using DFT within the local density approximation (LDA),\cite{PZ1981,PW1992} along with a plane wave basis set and norm-conserving pseudopotentials\cite{TM1991}, as implemented in the Quantum Espresso software suite.\cite{QE2009} A non-linear core correction is added to the In pseudopotential to treat the core-valence interaction.\cite{LouieFroyenCohen1982} 50 Rydberg kinetic energy cutoff is used for the plane wave basis set. Numerically converged Monkhorst-Pack\cite{MonkhorstPack1976} $k$-point grids
%%of $8\times8\times8$ and $4\times8\times1$
are used for all supercells in this work. The macroscopic average along the $z$ axis (aligned to the (001) direction) of the planar average (parallel to the interfacial plane) of the self-consistent potential\cite{Baldereschi88} [$V$$^{m}$$(z)$] for bulk and interface cells are calculated within DFT.\cite{Giantomassi11} Interface models consist of 24 atomic layers oriented along the (001) direction; 11 monolayers for Ge, 11 monolayers for In$_{x}$Al$_{1-x}$As, and at least 1 mixed monolayer per periodic image of the supercell. Parallel to the interface, interface supercells have dimensions of ($2\times1$) in units of the (110) lattice parameter. The virtual crystal approximation (VCA) is used to approximate the In$_{x}$Al$_{1-x}$As cation alloy for each composition point.  

Bulk cells are used to calculate the bulk band edges relative to the respective $V$$^{m}$$(z)$ for each material, and interface cells are used to calculate the potential offset ($dV$, see below) between the slabs. All band offsets correspond to fully relaxed geometries for bulk In$_{x}$Al$_{1-x}$As and interface cells, while for Ge the bulk cells are biaxially strained along the (100) and (010) directions and allowed to relax along (001). Thus, the bulk cells represent biaxially tensile strained Ge grown on an In$_{x}$Al$_{1-x}$As substrate (with AlAs lattice matched to Ge) and the interface models represent the minimum energy bonding configuration between the slabs. 

In order to investigate the relative stability of diffused impurities in either Ge or AlAs, large cubic bulk cells with dimensions (3$\times$3$\times$3) in units of the (100) lattice parameter are used to calculate the formation energies\cite{VandewalNeugebauer04,Freysholdt14} of substitutional impurities in bulk Ge and bulk AlAs. \footnote{We do not consider In$_{x}$Al$_{1-x}$As with $x$ $>$ 0 for this purpose due to the inaccurate bond lengths resulting from the VCA approach which would lead to inaccurate impurity-host bonding energies.} These correspond to impurity defects present after growth as a result of diffusion of substrate species during growth of the material on the substrate. Thus, we consider a single Al (As) on a Ge site (Al(As)$_\text{Ge}$) in a 216 atom bulk Ge supercell, and a single Ge on an Al (As) site (Ge$_\text{Al(As)}$) in a 216 atom bulk AlAs cell. The formation energies are calculated as a function of the chemical potential $\mu$$_\alpha$ of each exchanged atom $\alpha$\cite{VandewalNeugebauer04,Freysholdt14} which are related to the bulk elemental phases to establish boundaries on $\mu$$_\alpha$. Thus, the formation energies can be calculated for As-rich and Al-rich conditions, where the range of variation of As and Al chemical potentials corresponds to the heat of formation of AlAs\cite{ZhangNorthrup91}.

Thermodynamically stable configurations correspond to charge neutral interfacial bonding configurations between group-IV and group-III/V atoms\cite{Peressi98,Martin80,Kunc81,BylanderKleinman90} with no electric field building up across either material. In supercell simulations, the lack of a slab dipole is ensured when N$_{IV-V}$ = N$_{IV-III}$, where N$_{IV-V(III)}$ is the number of Ge-V(Ge-III) bonds per simulation cell. This constraint is imposed in all simulations in this work. 

Clearly, accurate band offsets require accurate calculations of the bulk band structures, which is precluded in DFT due to the well-known band gap problem.\cite{PerdewLevy82,ShamSchluter85,GodbySchluterSham88} The DFT+$GW$ approach corrects the energy levels using the $GW$ approximation to the electron self-energy\cite{Hedin1965,AryasetGunnars97,HybLouie86} providing sufficiently accurate bulk band structures for evaluating band alignments at semiconductor/oxide interfaces.\cite{Myrta10,Giantomassi11} 
% Accurate assessments of the band alignments require corrections from the GW method\cite{Hedin1965,AryasetGunnars97} for both VBOs and CBOs for these systems. 
In this work, differences of 0.17 eV between the VBO calculated with and without the $GW$ correction are found for the lattice matched Ge/AlAs(001) and for lattice mismatched $\varepsilon$-Ge/In$_{x}$Al$_{1-x}$As(001). For these reasons, all VBOs and CBOs calculated in this work are obtained using the DFT+$GW$ approach. This yields a first-order approximation to the quasiparticle band gaps from which band offsets are derived.\cite{Myrta10,Giantomassi11} Valence and conduction band offsets are computed using 
\begin{equation} \label{eq.1}
\Delta E_V = E_{V,Ge} - E_{V,III-V} + \Delta(\delta E_V) + dV \
\end{equation}
\begin{equation} \label{eq.2}
\Delta E_C = E_{C,III-V} - E_{C,Ge} - \Delta(\delta E_C) - dV \ 
\end{equation}
where $E_{V,Ge}$ ($E_{V,III-V}$) is the valence band maximum of Ge (In$_{x}$Al$_{1-x}$As) relative to $V^{m}$$(z)$ of the bulk cells, $E_{C,Ge}$ ($E_{C,III-V}$) is the DFT conduction band minimum of Ge (In$_{x}$Al$_{1-x}$As) relative to $E_{V,Ge}$ ($E_{V,III-V}$), $\delta$$E_V$ ($\delta$$E_C$) is the \textit{GW} correction to the valence band maximum (conduction band minimum) and $\Delta$($\delta$$E_{V/C}$) represents the difference between the materials in the \textit{GW} correction for the valence/conduction band edge (V/CBE). $dV$ is the offset in $V$$^{m}$$(z)$ across the interface. To obtain $dV$, the entire self-consistent potential was taken, however the change in $V^{m}$$(z)$ across the interface does not significantly involve the exchange-correlation potential, which is flat throughout the interface cell. For unstrained Ge, the conduction band minimum resides at the L point, while for sufficient biaxial strain $\varepsilon$$_{Ge}$ Ge exhibits a direct minimum gap. After relaxing the lattice constants of AlAs and InAs, and assuming a linear variation of the In$_{x}$Al$_{1-x}$As lattice constant with $x$, the corresponding change in cell parameters is applied to the relaxed Ge cell, resulting in $\varepsilon$$_{Ge}$ = 1.76\% when In content $x$ = 0.25. As discussed in the results section, the varying In content affects the ordering of the satellite valleys of both Ge and In, which has important implications for the band offsets. 

We note that no spin-orbit coupling (SOC) is included in these calculations. The effect of SOC is to split the heavy hole $p$ states (with $J$ = 1/2 angular momentum) near the top of the valence band. This splitting is 0.30 eV in Ge, and 0.275 eV in AlAs\cite{madelung2004}. As this difference of 0.025 eV is relatively small, we expect a correspondingly small effect on our calculated band offsets, which in this work always correspond to the offset between band extrema. Hence, we consider the gain in accuracy to be insufficient to justify the increased computational load of including relativistic terms, and we omit SOC.  

\begin{figure}[t]
\includegraphics[width=1.0\columnwidth]{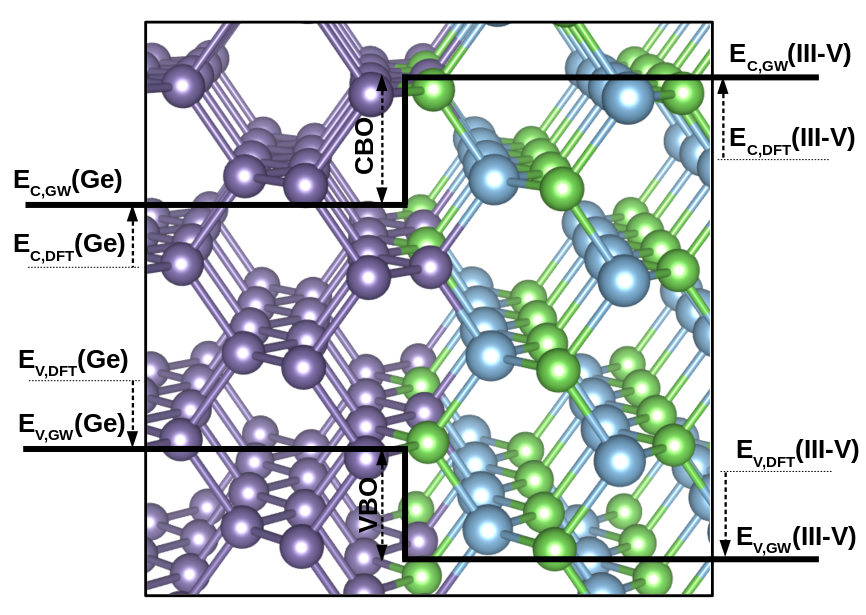}
\caption{\label{fig:interface_BOlines} Atomic structure of the ideal, ordered, As-terminated Ge / In$_{x}$Al$_{1-x}$As interface. Black lines correspond to a schematic representation of the valence and conduction BOs (with and without $GW$ corrections) across the interface. Ge atoms are purple, As atoms are green, and In/Al atoms are blue.}
\end{figure}

\section{Results} \label{results}
\subsection{Abrupt ordered interfaces} \label{sec-abrupt-int}

In this section, we focus on the interface that is atomically abrupt and localized to a single mixed monolayer which resides precisely between the slabs forming the heterojunction. The mixed interfacial monolayer (MIML) consists of either Ge and As atoms, or of Ge and In/Al atoms. This is exemplified by the ordered interface shown in Fig.~\ref{fig:interface_BOlines}. No interdiffusion is considered at this stage. Considering this as a fixed interfacial configuration, the In$_{x}$Al$_{1-x}$ stoichiometry of the substrate is varied from $x$ = 0 to $x$ = 0.25 and the valence and conduction band offsets are tracked in steps of $\Delta$$x$ = 0.05. Varying the In content affects the lattice constant of the substrate which in turn changes the strain state of the Ge slab. As $\varepsilon$$_{Ge}$ increases beyond 1.5\%, the conduction band satellite valleys are reordered in energy and Ge becomes a direct gap material. An analogous statement can be made for the conduction band valleys of In$_x$Al$_{1-x}$As. Our calculations show that for $x$ $\leq$ 0.20, the minimum energy valley in In$_x$Al$_{1-x}$As resides at the X point, while for larger proportions of In, In$_x$Al$_{1-x}$As exhibits a direct minimum gap at $\Gamma$.

\begin{figure}[t]
\includegraphics[width=\columnwidth]{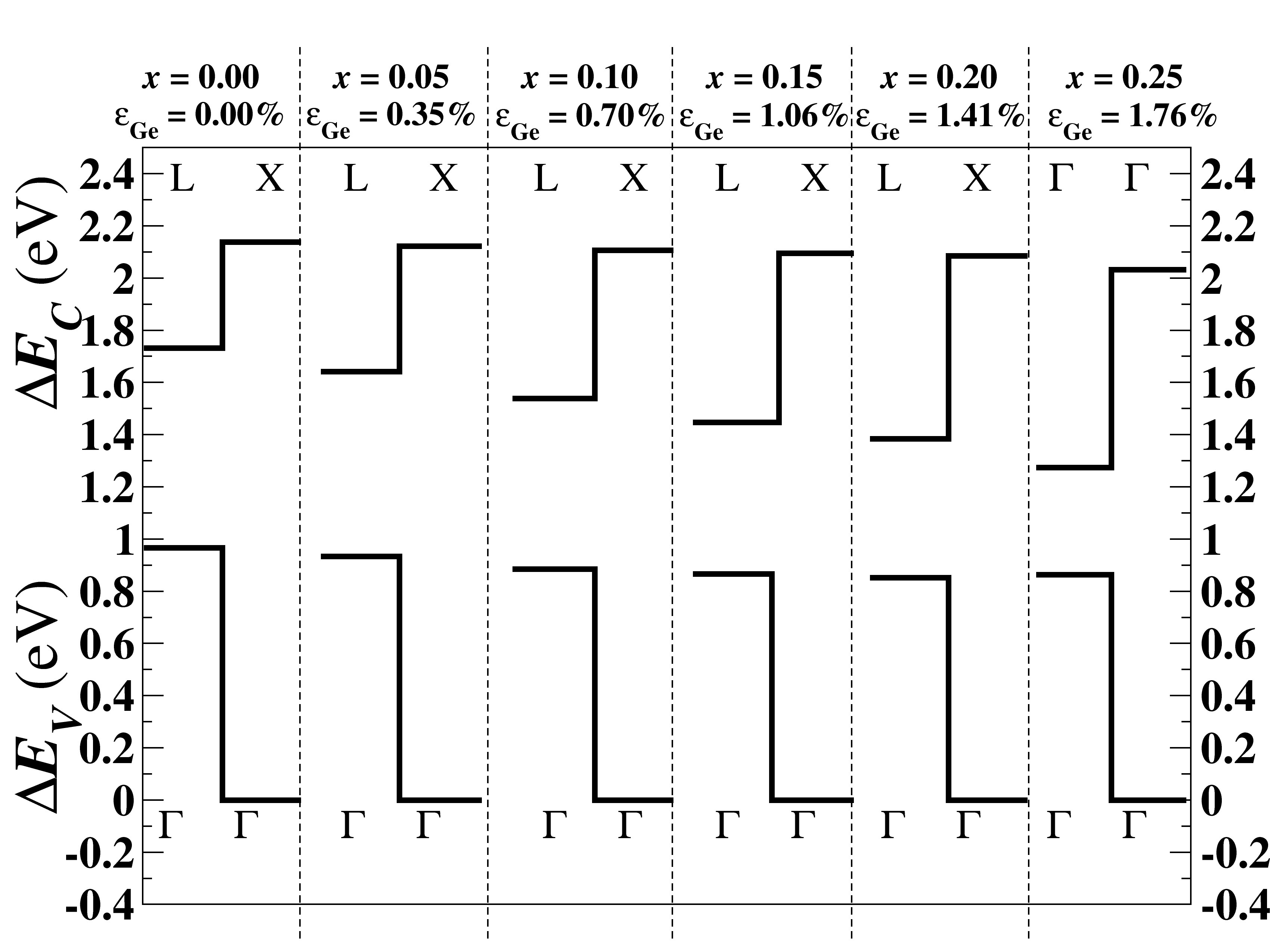}
\caption{\label{fig:VBOs_CBOs_InxAl1-xAs_As-Ge-int} Valence and conduction band offsets calculated using DFT+$GW$ for the abrupt As-terminated $\varepsilon$-Ge/In$_x$Al$_{1-x}$As(001) interface. The In content is varied in steps of 0.05, up to 0.25 (where 0.00 corresponds to AlAs). Variations of In content lead to corresponding changes in Ge strain, denoted by $\varepsilon$$_{Ge}$. The band gaps are labelled by the satellite valleys of minimum (maximum) energy for the conduction (valence) bands for both materials; for $x$ $\leq$ 0.20, the minimum energy conduction valley is L for Ge and X for In$_x$Al$_{1-x}$As.}
\end{figure}

In recent experimental works, III-V substrate growth is immediately followed by cooling under an As$_{2}$ overpressure before transfer to a vacuum chamber for Ge growth,\cite{Clavel2015} such that the resulting heterostructure most likely corresponds to Ge grown on an As-terminated In$_{x}$Al$_{1-x}$ slab, rendering an interfacial layer consisting of Ge and As atoms. In other experimental studies, group-III precursors was introduced immediately prior to Ge growth,\cite{Cheng2012} or even coverages of group-III and V atoms on the III-V surface before Ge growth were inferred from the observed surface reconstruction\cite{Maeda1995} with significant III segregation into Ge after Ge growth.\cite{Maeda1999} All of these studies taken together provide an impetus to study both III-terminated and V-terminated In$_{x}$Al$_{1-x}$As interfaced with Ge. As will be shown below, alternative interfacial stoichiometries can lead to interesting behavior in the form of qualitative changes to the band alignments. Hence, we study BOs for both cases, starting with the interface in which In$_x$Al$_{1-x}$As is As terminated (see Fig.~\ref{fig:interface_BOlines} for structure). The results are displayed in Fig.~\ref{fig:VBOs_CBOs_InxAl1-xAs_As-Ge-int}. 

With the interfacial configuration fixed to that shown in Fig.~\ref{fig:interface_BOlines} (group-V-terminated), a relatively small change in the valence band alignment is observed as a function of In content; 0.11 eV change in the VBO is observed between $x$ = 0.00 and $x$ = 0.25. The CBO exhibits a larger change of 0.47 eV as a function of In content. Thus a type-I BO is observed for Ge interfaced with As-terminated In$_x$Al$_{1-x}$As, and modifications of the group-III composition of the substrate for 0.00 $\leq$ $x$ $\leq$ 0.25 (neglecting changes due to the randomized cation alloy) does not qualitatively change the band alignment.

\begin{figure}[t]
\includegraphics[width=\columnwidth]{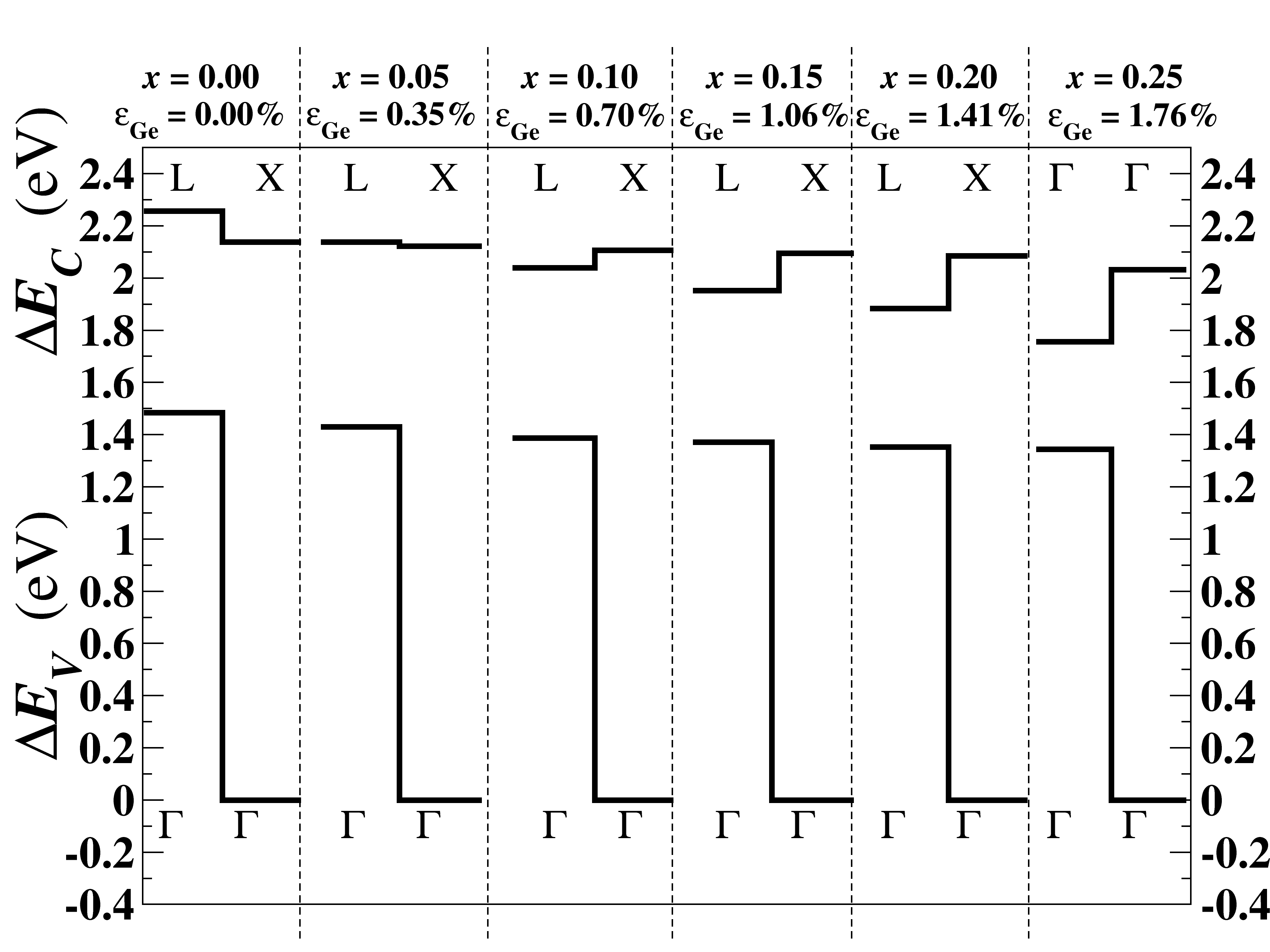}
\caption{\label{fig:VBOs_CBOs_InxAl1-xAs_III-Ge-int} Valence and conduction band offsets calculated using DFT+$GW$ for the abrupt III-terminated $\varepsilon$-Ge/In$_x$Al$_{1-x}$As interface. The In content is varied in steps of 0.05, up to 0.25 (where 0.00 corresponds to AlAs). Variations of In content lead to corresponding changes in Ge strain, denoted by $\varepsilon$$_{Ge}$. The band gaps are labelled by the satellite valleys of minimum (maximum) energy for the conduction (valence) bands for both materials; for $x$ $\leq$ 0.20, the minimum energy conduction valley is L for Ge and X for In$_x$Al$_{1-x}$As.}
\end{figure}

Changing to the group-III-terminated In$_x$Al$_{1-x}$As, in which the MIML consists of Ge and In/Al cation atoms, a stark contrast is observed in the band alignments compared to the As-terminated case. Fig.~\ref{fig:VBOs_CBOs_InxAl1-xAs_III-Ge-int} shows much larger VBOs and correspondingly smaller CBOs, with the CBO becoming negative (corresponding to the Ge CBE being higher in energy than that of In$_x$Al$_{1-x}$As, see equation (\ref{eq.2})) for small values of $x$; a type-II band offset is calculated for $x$ $<$ 0.05, and so a type-II to type-I transition in the band alignment occurs as a function of III content for the abrupt III-terminated $\varepsilon$-Ge/In$_x$Al$_{1-x}$(001) interface, with the BOs being type-I for $x$ $>$ 0.05. 

In addition to showing the BO dependence on substrate stoichiometry, this is also a strong indication of the high sensitivity of band alignments to interfacial stoichiometry for $\varepsilon$-Ge/In$_{x}$Al$_{1-x}$As(001) (compare Figs.~\ref{fig:VBOs_CBOs_InxAl1-xAs_As-Ge-int} and ~\ref{fig:VBOs_CBOs_InxAl1-xAs_III-Ge-int} for a given value of $x$), and shows that the band alignment can also change from type-I to type-II when comparing As-terminated to III-terminated Ge/AlAs interfaces. This is in qualitative agreement with the results of Pavarelli {\it et al}.\cite{Pavarelli13} who reported an analogous change in the calculated band offset type for anion and cation dominated interface stoichiometries in the $\varepsilon$-Ge/In$_{0.3}$Ga$_{0.7}$As(001) interface. The results also indicate a comparable change in the VBO and CBO as a function of In content for both the III-terminated case and the As-terminated case, although a slightly larger change is seen for the III-terminated case. This is explained by the presence of group-III atoms at the interface. As the latter corresponds to a III rich In$_x$Al$_{1-x}$As surface, the larger variation of the VBO (0.15 eV) and CBO (0.5 eV) with respect to the In$_x$Al$_{1-x}$ stoichiometry is observed due to the slightly larger effect that $x$ has on the interface potential term $dV$, where the latter is derived from the atomic potentials present in the interface supercell.

\subsection{Interdiffusion} \label{sec-intdiff}

\begin{figure} [t]
\includegraphics[width=10cm,height=3.5cm]{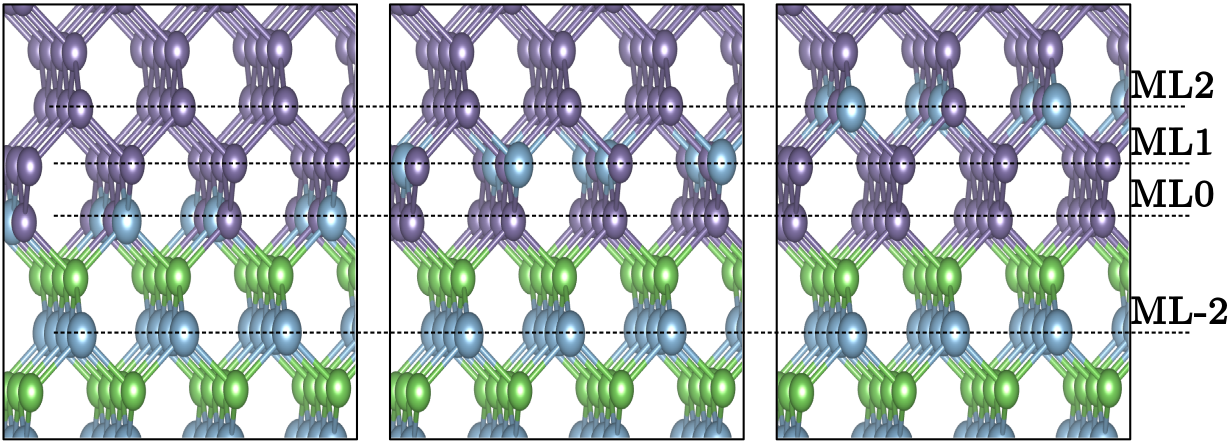}
\caption{\label{fig:GeOnAlAs_for_paper_ML012-2_AlGe} (Left) Model of MIML residing exactly between the materials forming the heterojunction; this position is labelled ML0. (Middle) MIML positioned one monolayer further from ML0, towards the Ge slab, with position labeled ML1. (Right) MIML positioned 2 monolayers towards the Ge slab, with position labeled ML2. For Ge diffusing into In$_{x}$Al$_{1-x}$As, we consider Ge atoms residing up to two monolayers away from the ML0 position, towards In$_{x}$Al$_{1-x}$As, with position labeled ML-2. Ge atoms are purple, group-V As atoms are green, and group-III In and Al atoms are blue. In this figure, the case of group-III atoms residing in the top In$_{x}$Al$_{1-x}$As layer is used as an example. For the cases of As atoms residing in the top layer, all green and blue atoms are exchanged.}
\end{figure}

In order to investigate the effects of interdiffusion of atomic species across the interface on band offsets, the position of the MIML was shifted up to 2 monolayers away from the abrupt interfacial layer separating the materials, either towards Ge or towards In$_x$Al$_{1-x}$As (Fig.~\ref{fig:GeOnAlAs_for_paper_ML012-2_AlGe}).  This corresponds to a maximum thickness of $\sim$6 {{\AA}} over which atomic diffusion is considered (i.e. $\pm$$\sim$3 {{\AA}} from the ML0 position, see Fig.~\ref{fig:GeOnAlAs_for_paper_ML012-2_AlGe}), which is consistent with previous experimental reports of interface abruptness in comparable heterostructures.\cite{Clavel2015,Nguyen2015} However, these interfacial configurations are unrealistic, and for heterostructures present in experimental samples, interfacial configurations involving mixed depths of diffusing species throughout the interfacial region are much more likely to occur. As a first approximation, this can be investigated by linearly varying the stoichiometric balance of atoms between adjacent MIMLs (while always maintaining charge neutral configurations) near the interface.

For example, consider the Ge/AlAs heterojunction with an abrupt interface in which the MIML consists of Ge and Al atoms (corresponding to the band offsets on the far left for $x$ = 0 and $\varepsilon$$_{Ge}$ = 0 in Fig.~\ref{fig:VBOs_CBOs_InxAl1-xAs_III-Ge-int}). This position of the MIML is referred to as ML0 (see right panel of Fig.~\ref{fig:GeOnAlAs_for_paper_ML012-2_AlGe}). By using the VCA to linearly mix the atoms of ML0 and ML1 (see middle panel of Fig.~\ref{fig:GeOnAlAs_for_paper_ML012-2_AlGe}), an approximation to an interfacial configuration involving mixed diffusion depths of Al atoms into the Ge slab can be achieved. For the case of Al atoms diffusing from ML0 to ML1, the stoichiometric balance between the monolayers required to maintain neutrality results in the relation 

\vspace{5mm}

\centerline{[Al$_{0.5-−a}$Ge$_{0.5+a}$]$^\textrm{ML0}$ = [Al$_{a}$Ge$_{1−-a}$]$^\textrm{ML1}$}

\vspace{5mm}

\noindent where [Al$_{a}$Ge$_{1-a}$]$^\textrm{ML0/1}$ is the composition of ML0/1, and $a$ is varied from 0 to 0.5. This is repeated for the case of Al atoms diffusing between ML1 and ML2, with $b$ used as the stoichiometry parameter instead of $a$ to avoid confusion. For the case of As atoms diffusing into Ge (see Sec.~\ref{sec-Asdiff}), the stoichiometric relation is analogous, with Al sites being replaced by As. This procedure is also repeated for the case of Ge atoms diffusing into In$_x$Al$_{1-x}$As (see Sec.~\ref{sec-Gediff}) which would more likely correspond to the scenario of a III-V slab grown on a Ge substrate.\cite{Chia2008} The stoichiometries of the endpoints (e.g. Al$_{0.5}$Ge$_{0.5}$ in ML0, ML1, or ML2, see Fig.~\ref{fig:GeOnAlAs_for_paper_ML012-2_AlGe}) are calculated using explicit atomistic models (see Fig.~\ref{fig:VBOs_CBOs_InxAl1-xAs_III-Ge-int}) and compared to the corresponding VCA results for comparison. 

As an additional assessment of the error associated with modeling the mixed layer stoichiometries with the VCA, the cluster expansion formalism is used to generate a special quasirandom structure (SQS)\cite{Wei90} representation of the mixed monolayer with $a$ = 0.5. The ATAT code\cite{ATAT02,ATAT-mcsqs13} is used to generate the periodic monolayer cell which exhibits multisite correlation functions for $m$th nearest neighbor $k$-atom clusters $\bar{\Pi}_{k,m}^\text{2D}$ that match those of an infinite random 2 dimensional binary alloy with 50/50 composition, for up to $m$ = 2 and $k$ = 3. This monolayer SQS is used to compare to the ordered monolayer structures, and to the end-point VCA stoichiometries. This is done for Al diffusing into Ge, and for Ge diffusing into AlAs.

\begin{figure*}

{
\includegraphics[width=\columnwidth]{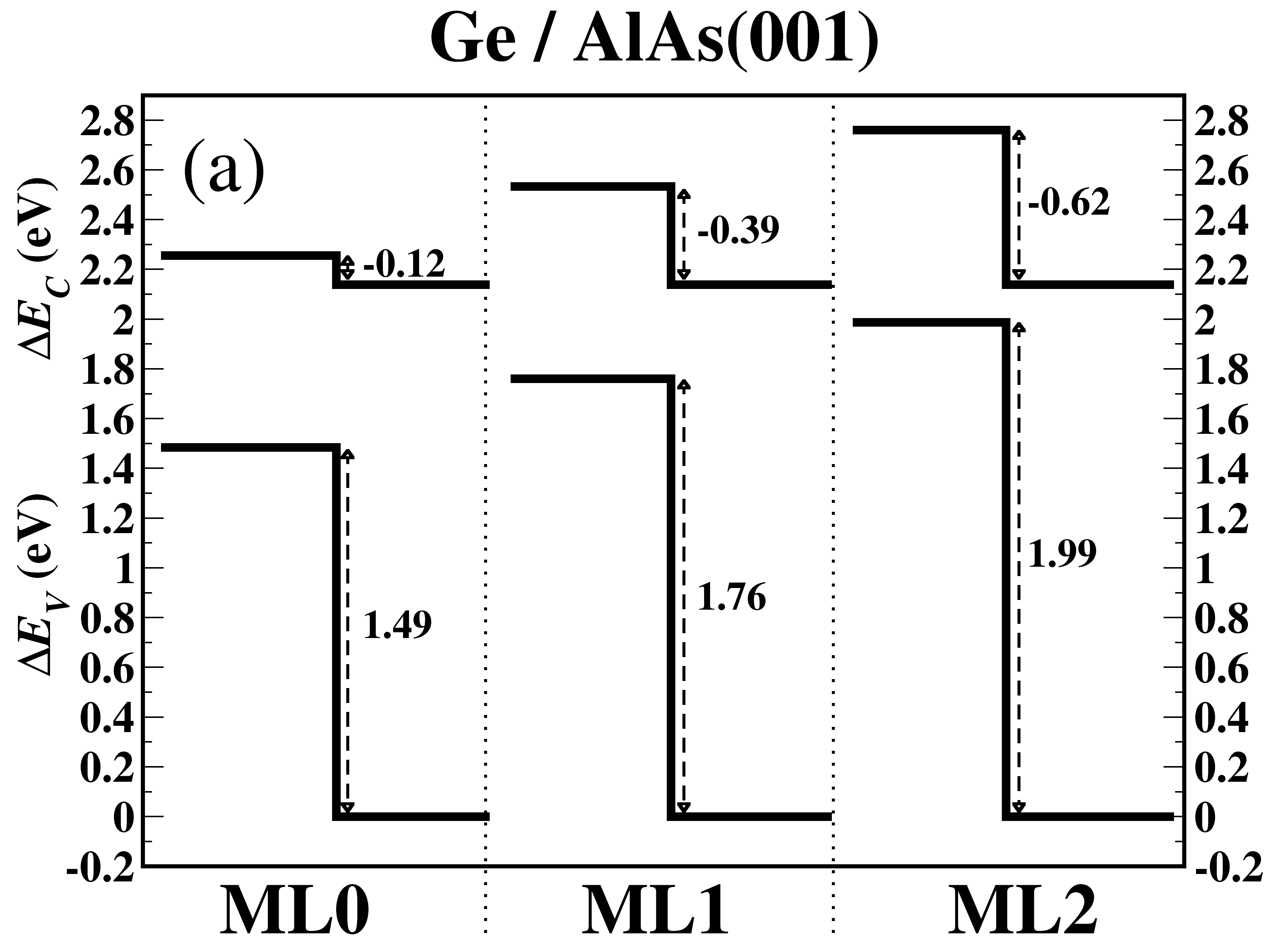} 
\includegraphics[width=\columnwidth]{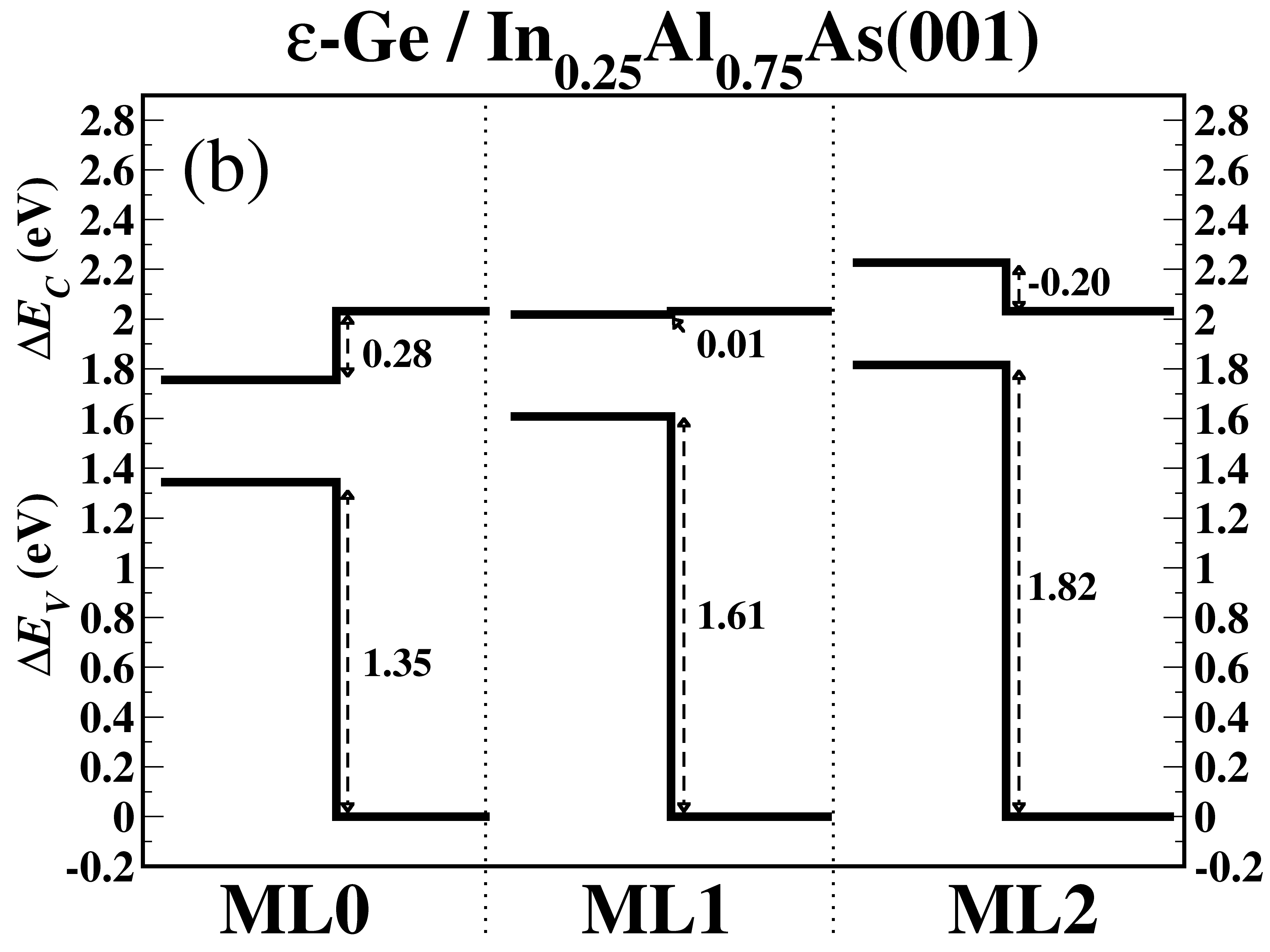}}

{
\includegraphics[width=\columnwidth]{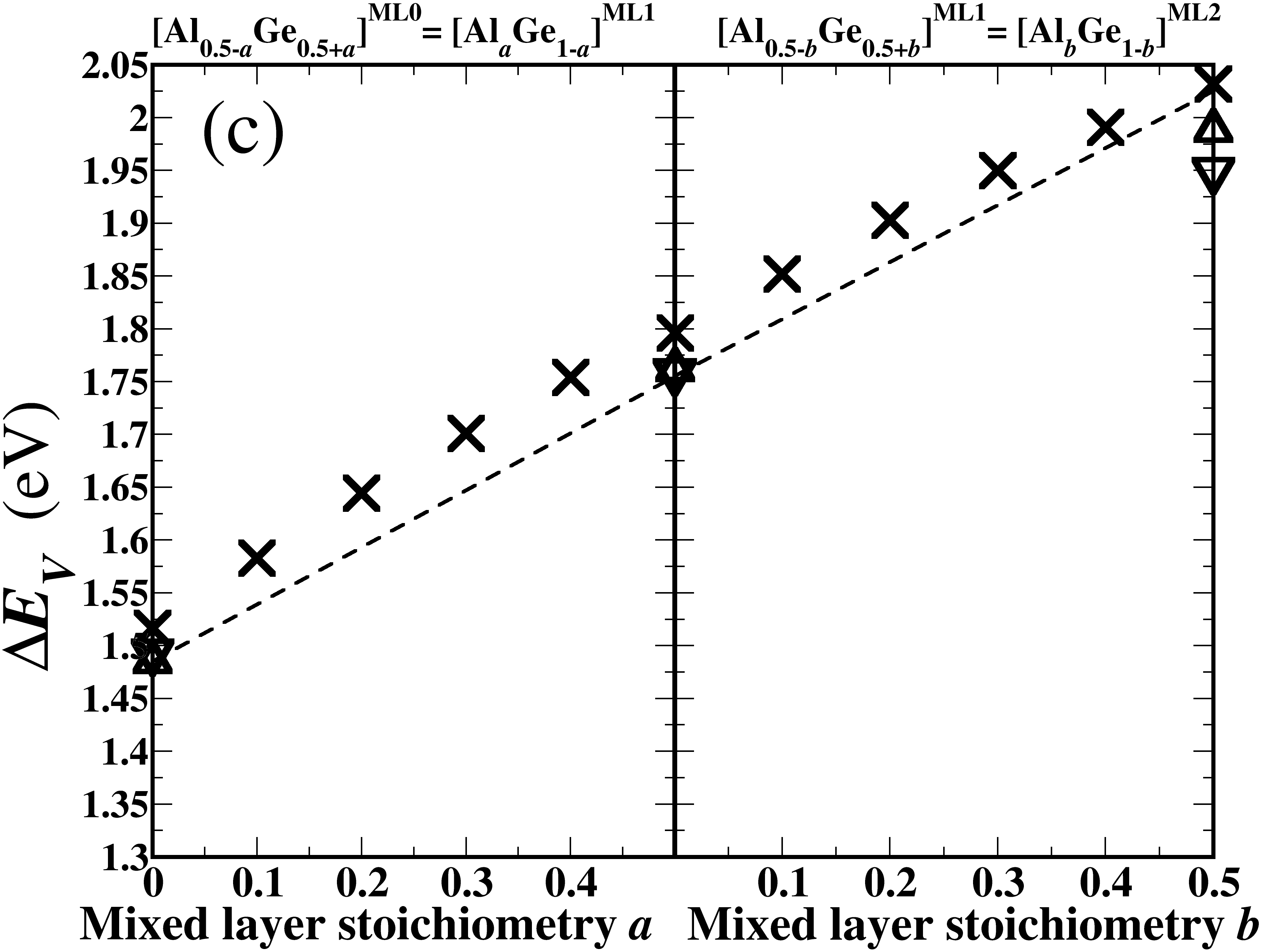}
\includegraphics[width=\columnwidth]{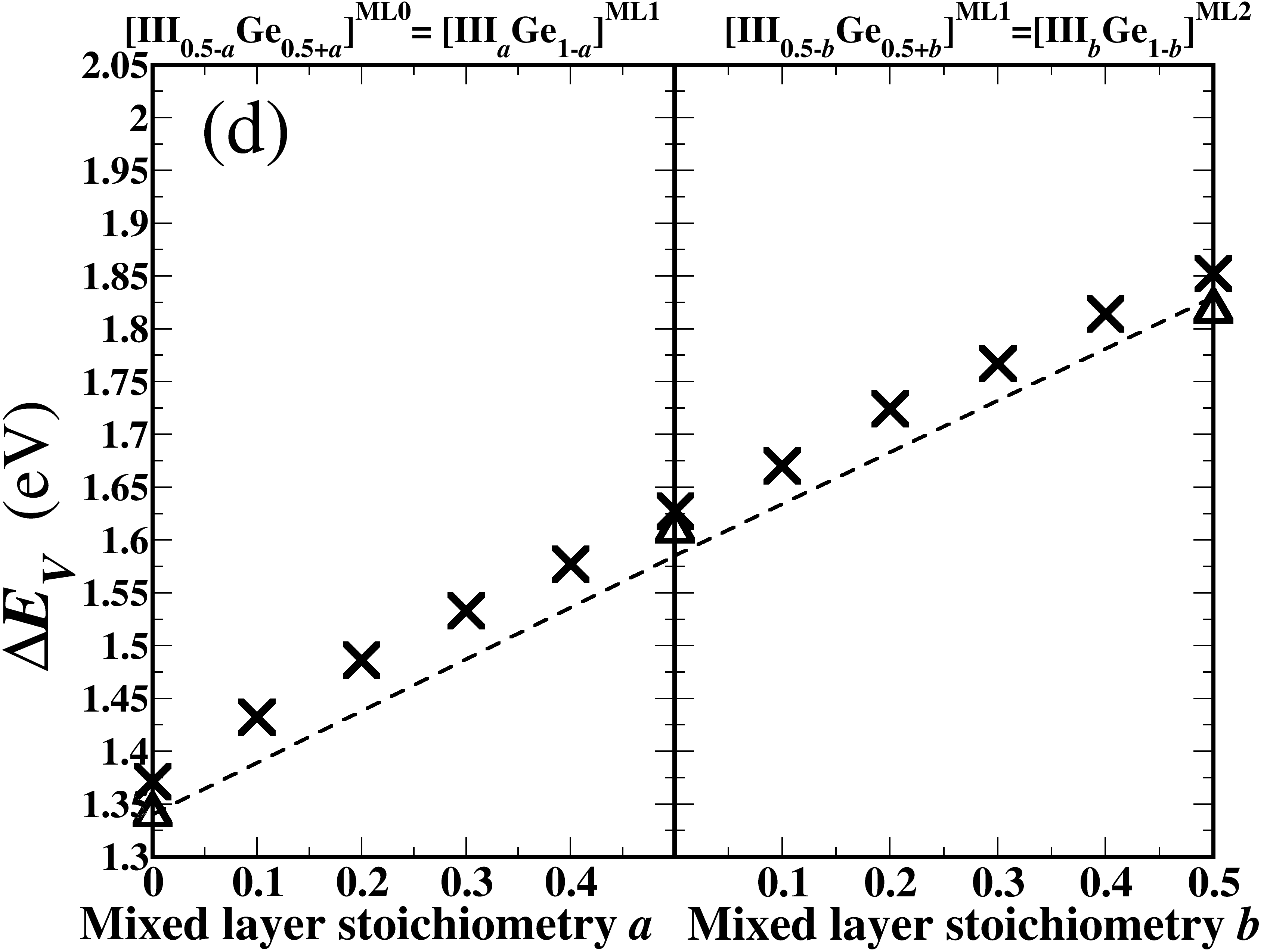}}

\caption[width=\columnwidth]{\label{fig:cat-diffusion_Ge-InxAl1-xAs}In (a) and (b), band offsets are presented for explicit (ordered) models of group-III In and Al atoms in the Ge slab, with increasing distance from the ML0 position. In (c) and (d), VBOs are plotted for group-III atoms diffusing into Ge, using the VCA to approximate the stoichiometry of monolayers near the interfacial plane. Left panels ((a) and (c)) correspond to Ge/AlAs(001), right panels ((b) and (d)) correspond to $\varepsilon$-Ge/In$_{0.25}$Al$_{0.75}$As(001). The \textbf{{\texttimes}} symbols label VBOs calculated using the VCA. The \textbf{$\triangle$} symbols correspond to explicit models of the atomic configurations for endpoint interface stoichiometries, ordered along the (110) direction parallel to the interface. The \textbf{$\bigtriangledown$} symbols represent the SQS model for the mixed monolayer. The dashed lines correspond to the linear response model (described in Sec.~\ref{sec-analysis}) for polar interfaces, applied to diffusion. As group-III In$_{x}$Al$_{1-x}$ cations diffuse away from the substrate into Ge, the band offsets become increasingly type-II in character for Ge/AlAs(001), and change from type-I to type-II for $\varepsilon$-Ge/In$_{x}$Al$_{1-x}$As(001). Note that negative values of the CBO correspond to the Ge CBE residing at a higher energy than the AlAs CBE.}
\end{figure*}

\begin{figure*}

{
\includegraphics[width=\columnwidth]{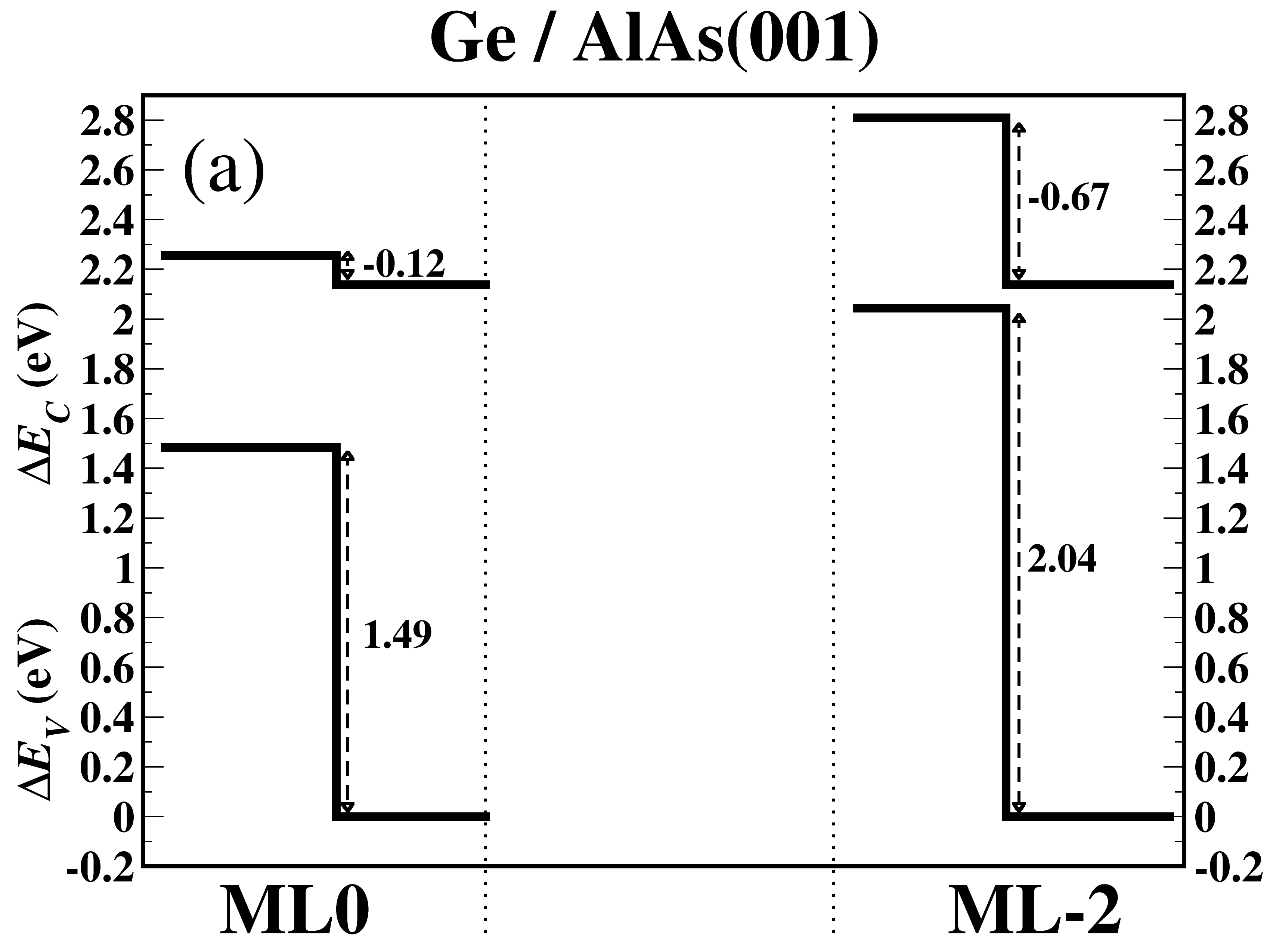} 
\includegraphics[width=\columnwidth]{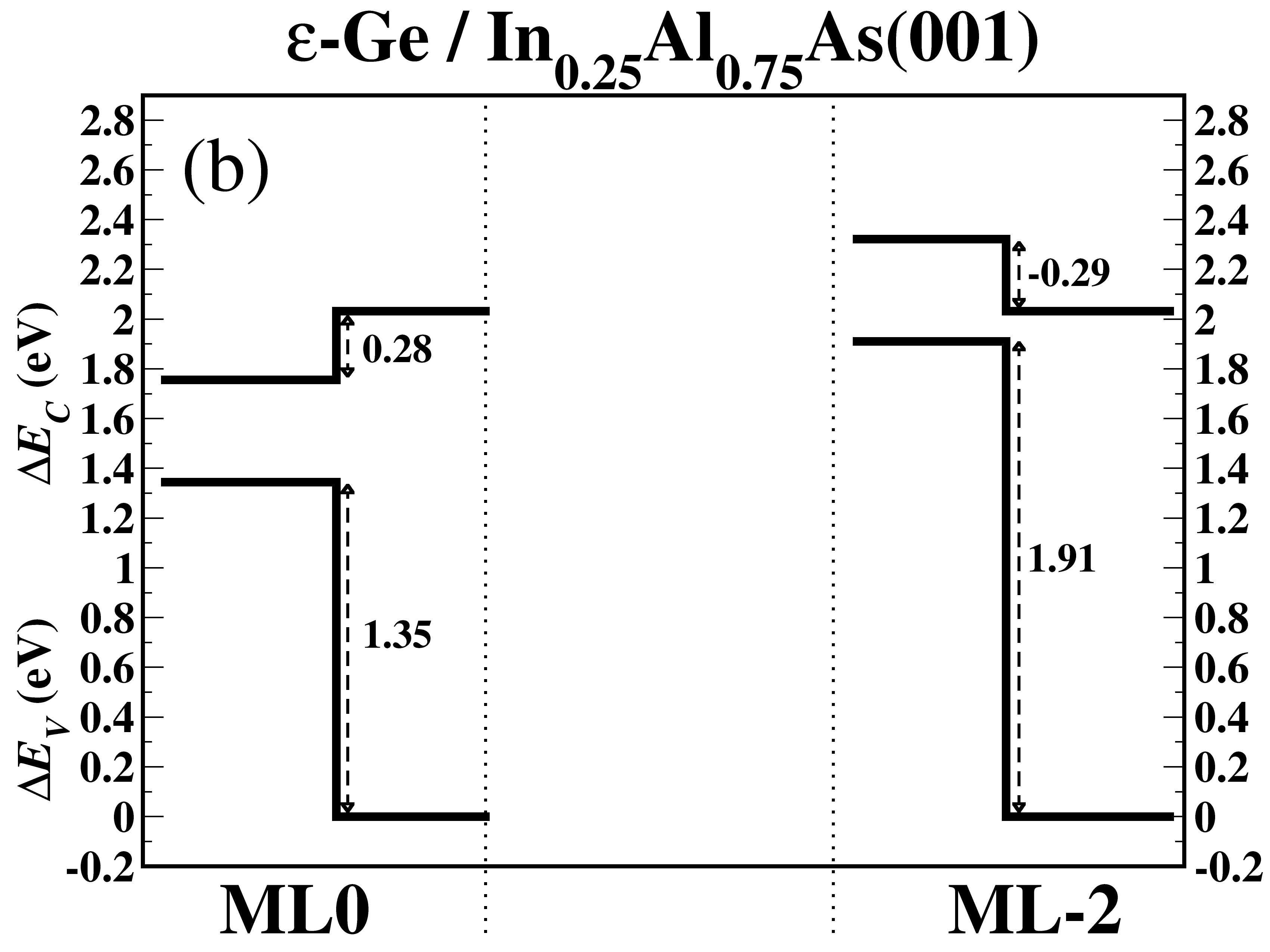}}

{
\includegraphics[width=\columnwidth]{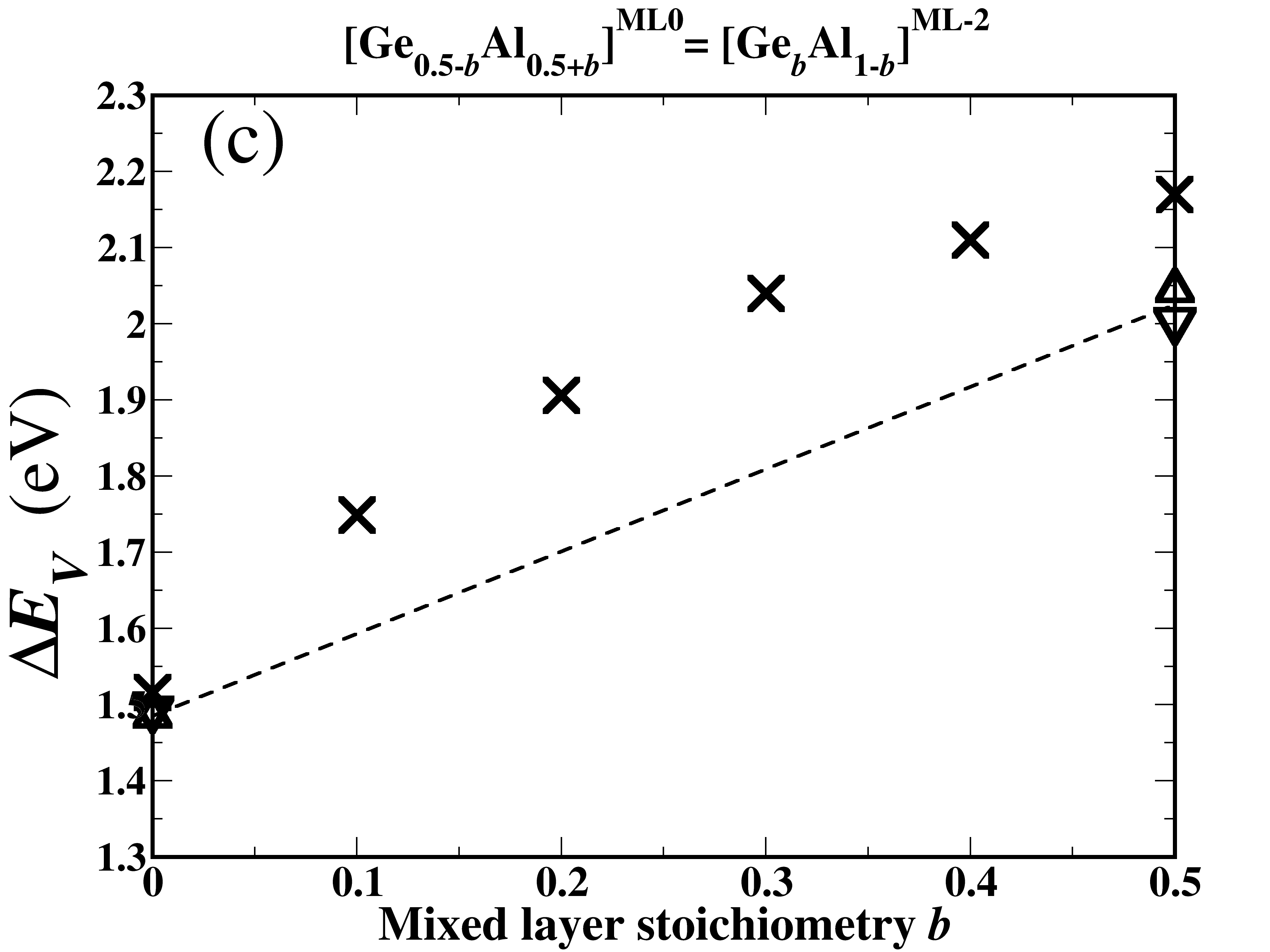}
\includegraphics[width=\columnwidth]{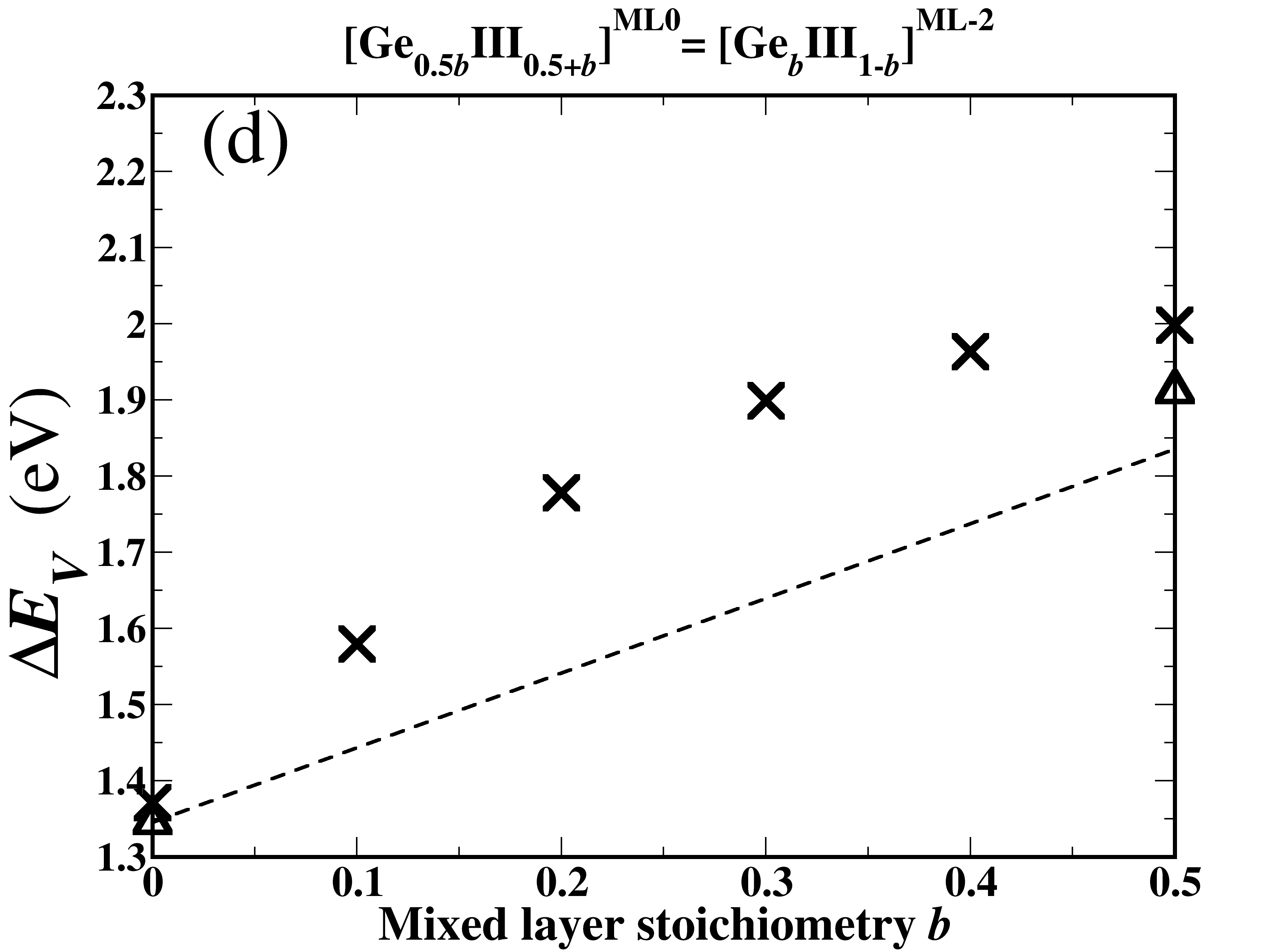}}

\caption[width=\columnwidth]{\label{fig:Ge-diffusion_GecatML_Ge-InxAl1-xAs}In (a) and (b), band offsets are presented for explicit models of Ge in a group-III layer of In$_{x}$Al$_{1-x}$As. In (c) and (d), VBOs are plotted for Ge diffusing from ML0 to ML-2 (the second group-III layer away from ML0), using the VCA to approximate the stoichiometry of monolayers near the interfacial plane. In$_{x}$Al$_{1-x}$As is III-terminated. Left panels ((a) and (c)) correspond to Ge/AlAs(001), and right panels ((b) and (d)) correspond to $\varepsilon$-Ge/In$_{0.25}$Al$_{0.75}$As(001). The \textbf{{\texttimes}} symbols label VBOs calculated using the VCA. The \textbf{$\triangle$} symbols correspond to explicit models of the atomic configurations for endpoint interface stoichiometries, ordered along the (110) direction parallel to the interface. The \textbf{$\bigtriangledown$} symbols represent the SQS model for the mixed monolayer. The dashed lines correspond to the linear response model (described in Sec.~\ref{sec-analysis}) for polar interfaces, applied to diffusion. For this range of diffusion, the band alignment remains type-II as a function of Ge diffusion distance into AlAs, and changes from type-I to type-II for Ge diffusion into In$_{0.25}$Al$_{0.75}$As.}
\end{figure*}

\begin{figure*}

{
\includegraphics[width=\columnwidth]{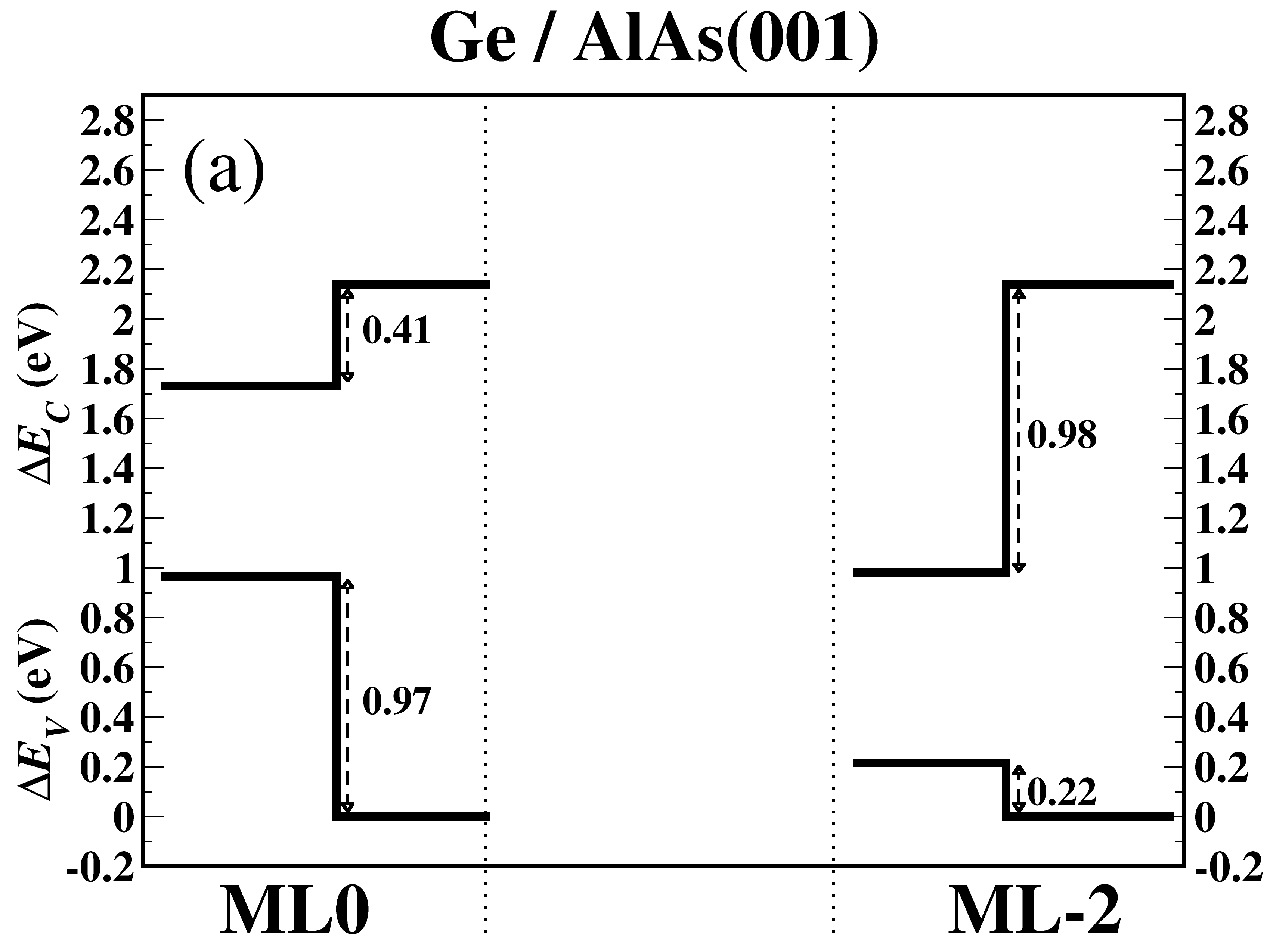} 
\includegraphics[width=\columnwidth]{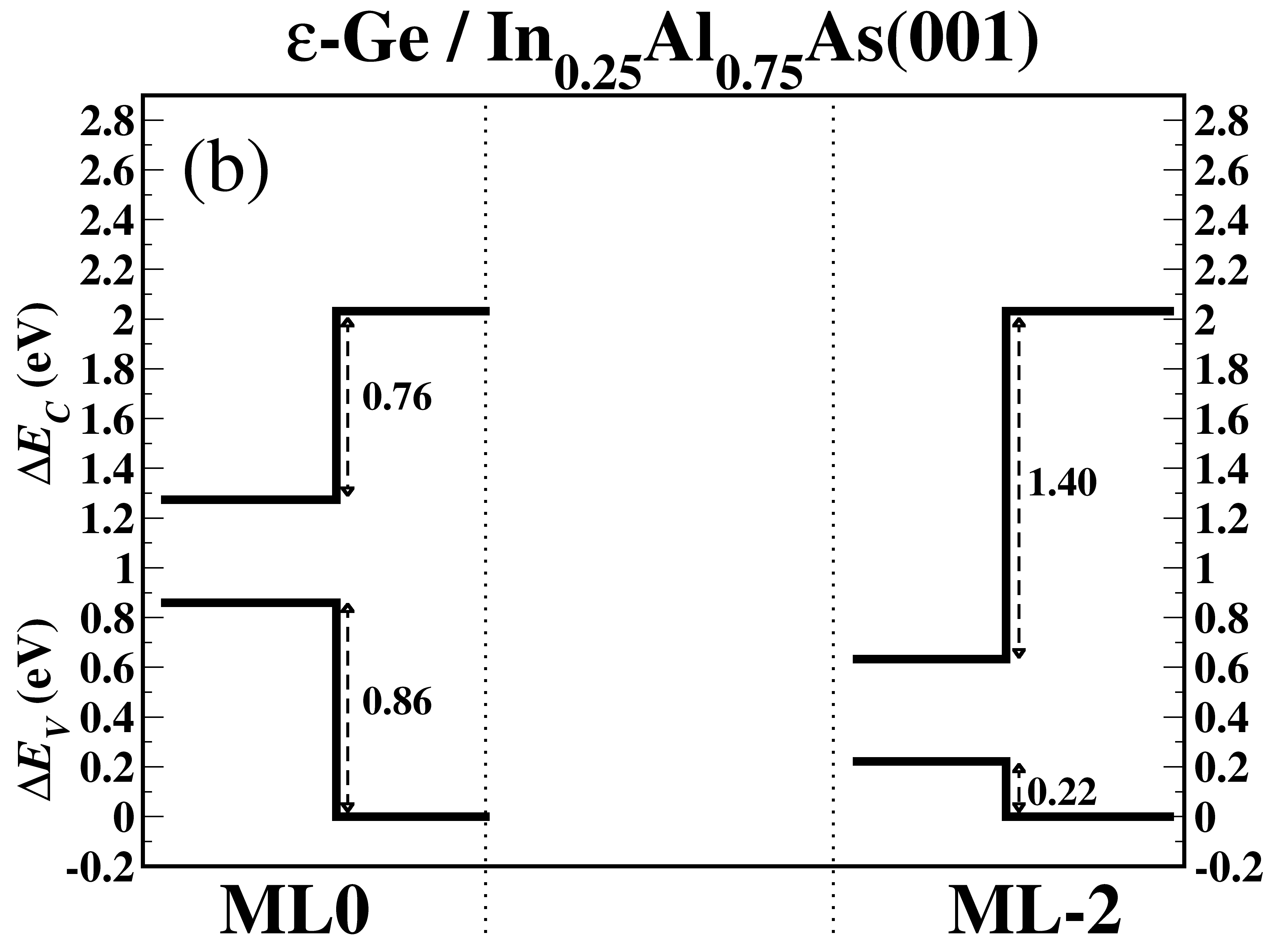}}

{
\includegraphics[width=\columnwidth]{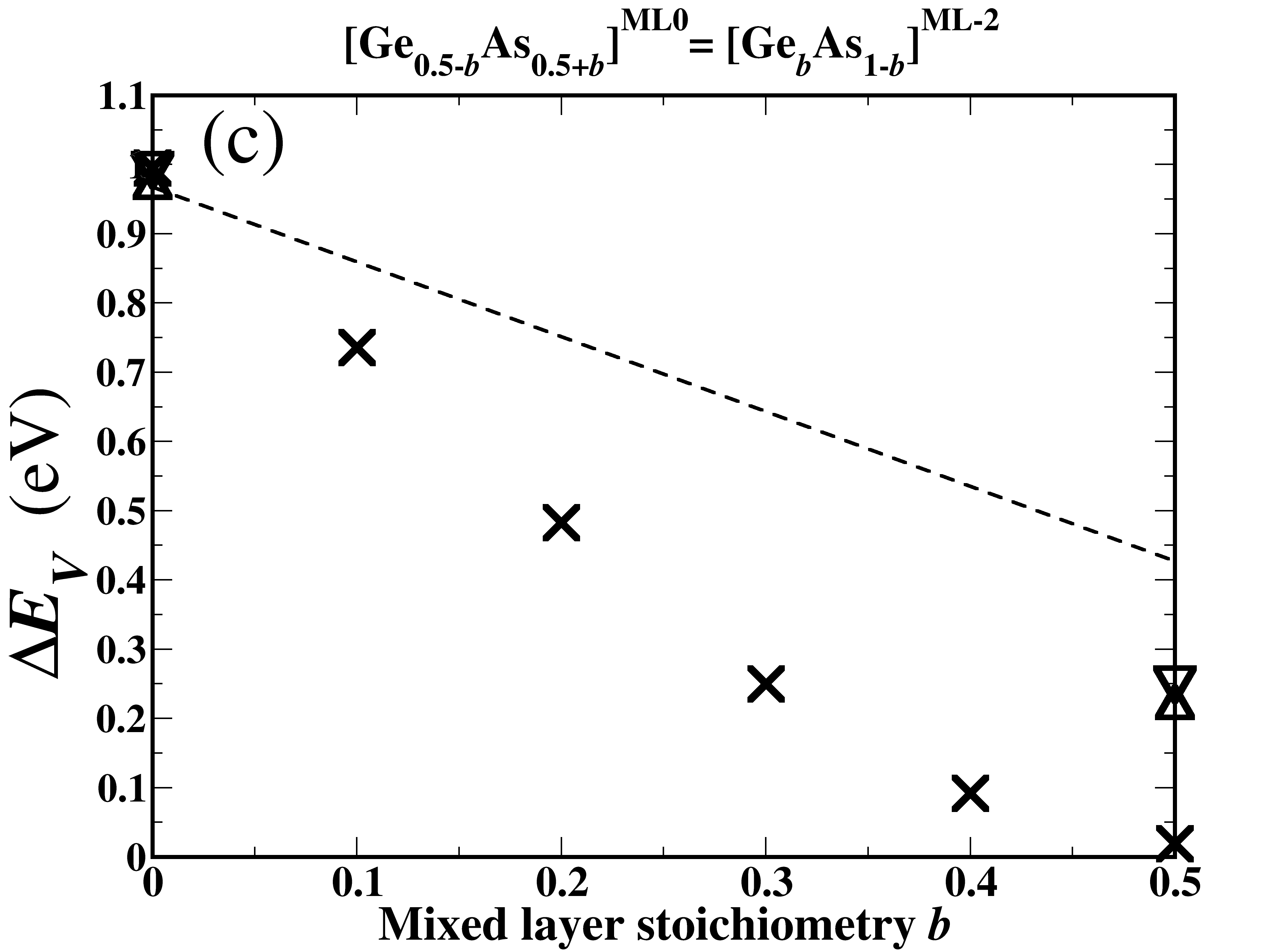}
\includegraphics[width=\columnwidth]{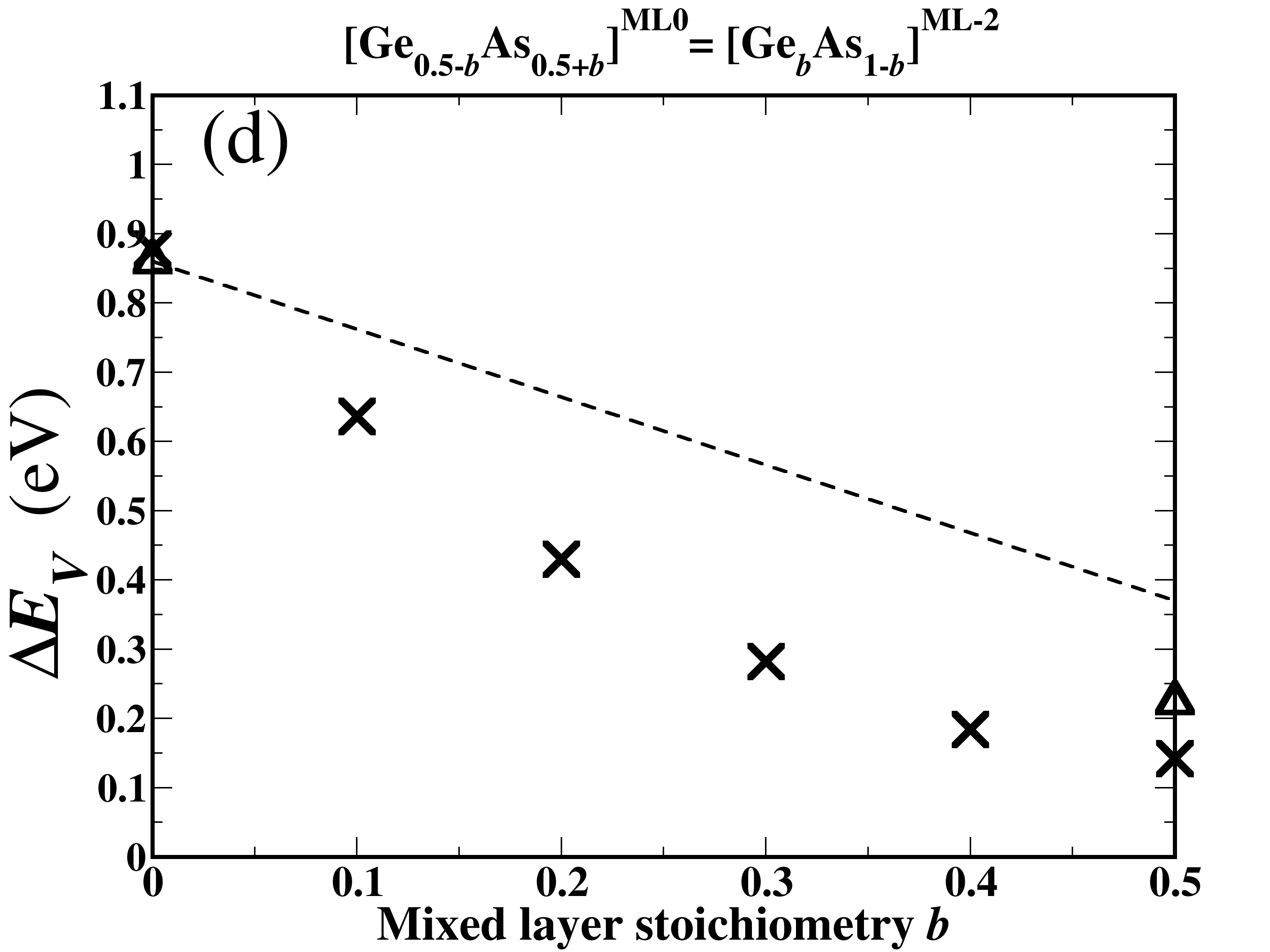}}

\caption[width=\columnwidth]{\label{fig:Ge-diffusion_GeAsML_Ge-InxAl1-xAs}In (a) and (b), band offsets are presented for explicit models of Ge in an As layer of In$_{x}$Al$_{1-x}$As. In (c) and (d), VBOs are plotted for Ge diffusing from ML0 to ML-2 (the second group As layer away from ML0), using the VCA to approximate the stoichiometry of monolayers near the interfacial plane. In$_{x}$Al$_{1-x}$As is As-terminated. Left panels ((a) and (c)) correspond to Ge/AlAs(001), and right panels ((b) and (d)) correspond to $\varepsilon$-Ge/In$_{0.25}$Al$_{0.75}$As(001). The \textbf{{\texttimes}} symbols label VBOs calculated using the VCA. The \textbf{$\triangle$} symbols correspond to explicit models of the atomic configurations for endpoint interface stoichiometries, ordered along the (110) direction parallel to the interface. The \textbf{$\bigtriangledown$} symbols represent the SQS model for the mixed monolayer. The dashed lines correspond to the linear response model (described in Sec.~\ref{sec-analysis}) for polar interfaces, applied to diffusion. For this range of diffusion, the band alignment remains type-I as a function of Ge diffusion distance into AlAs or In$_{0.25}$Al$_{0.75}$As, with the VBO significantly reduced in both cases.}
\end{figure*}

The results of these simulations are also compared to an analytical model based on a linear response theory for polar, heterovalent interfaces\cite{Peressi98,Harrison78}. This model is described in Sec.\ref{sec-analysis}, and the results obtained from this model are shown as dashed lines in Figs.~\ref{fig:cat-diffusion_Ge-InxAl1-xAs}, ~\ref{fig:Ge-diffusion_GecatML_Ge-InxAl1-xAs}, and ~\ref{fig:Ge-diffusion_GeAsML_Ge-InxAl1-xAs}. %in which the VBO is linearly extrapolated from the VBO obtained from abrupt (ML0), ordered models.

\begin{table}
\caption{\label{tab:AsGe_BO}
Calculated band offsets of Ge/AlAs(001) and $\varepsilon$-Ge/In$_{0.25}$Al$_{0.75}$As(001) for As-terminated In$_{x}$Al$_{1-x}$As, in which As atoms have diffused up to two monolayers into Ge. The MIML column refers to the position of the mixed interfacial monolayer, as defined by Fig.~\ref{fig:GeOnAlAs_for_paper_ML012-2_AlGe}. The values in brackets refer to estimates from the linear response model for polar interfaces\cite{Peressi98,Harrison78} applied to diffusion, as described in Sec.~\ref{sec-analysis}. All band offsets are in eV.}
\begin{ruledtabular}
  \begin{tabular}{c|cccc}
MIML & VBO & CBO & \\
\hline\\[-0.2cm]
 &Ge/AlAs(001) && \\[0.1cm]
ML0 & 0.97 & 0.41 & \\
ML1 & 0.74 (0.70) & 0.65 (0.61) & \\
ML2 & 0.46 (0.43) & 0.92 (0.89) & \\
\\[0.1cm]
& $\varepsilon$-Ge/In$_{0.25}$Al$_{0.75}$As(001) && \\[0.1cm]
ML0 & 0.86 &  0.75 &  \\
ML1 & 0.72 (0.62) &  0.90 (0.80) &  \\
ML2 & 0.48 (0.37) &  1.15 (1.04) &  \\
  \end{tabular}
\end{ruledtabular}
\end{table}

\subsubsection{In$_{x}$Al$_{1-x}$ diffusion into Ge} \label{sec-catdiff}

For the case of Al atoms diffusing away from the Ge/AlAs(001) interface and into Ge, a linear change in the band offset is observed (panels (a) and (c) of Fig.~\ref{fig:cat-diffusion_Ge-InxAl1-xAs}). The band alignment is type-II for the case of the MIML residing at ML0 ([Al$_{0.5}$Ge$_{0.5}$]$^\textrm{ML0}$). Thus, for Ge films grown on Al-terminated AlAs(001), the valence (conduction) band edge of AlAs resides above (below) that of Ge, and for increasing diffusion depth of Al atoms into Ge the CBO becomes increasingly negative and the VBO increasingly positive. As a result, the band alignment is increasingly type-II over this range as a function of diffusion distance of Al. While a diffusion distance of up to 2 monolayers (corresponding to $\sim$3 {{\AA}}) into Ge is particularly short, an increase in the band alignments of 0.50 eV (0.51 eV) is calculated for the explicit (VCA) models of the interface, which shows again the large sensitivity of band alignments to diffusion distance.

Turning to the lattice-mismatched interface $\varepsilon$-Ge/In$_{0.25}$Al$_{0.75}$As(001), a qualitatively similar movement of VBO and CBO with respect to diffusion distance of group-III (In and Al) atoms compared to Ge/AlAs(001) is calculated. The major difference compared to Ge/AlAs(001) is that the VBO for [(InAl)$_{0.5}$Ge$_{0.5}$]$^\textrm{ML0}$ is small enough to yield a type-I band alignment. A type-I to type-II transition in the band alignment is observed as a function of diffusion distance for this case, with the CBO for [(InAl)$_{0.5}$Ge$_{0.5}$]$^\textrm{ML1}$ being close to flat (0.01 eV) and [(InAl)$_{0.5}$Ge$_{0.5}$]$^\textrm{ML2}$ exhibiting a type-II band offset (Fig.~\ref{fig:cat-diffusion_Ge-InxAl1-xAs} (b)). 

Thus for group-III cations diffusing across the interface and into $\varepsilon$-Ge, calculations show that this can have a large enough effect as to change the character of the band alignment relative to the abrupt interface, even for a very short diffusion distance of two monolayers. From the perspective of device physics, this finding has significant consequences. For example, for devices involving sandwiches of $\varepsilon$-Ge between In$_{0.25}$Al$_{0.75}$As(001) layers, the trapping of both electrons and holes (required for optically active recombination in optoelectronic applications) will be highly dependent on the diffusion depth of In and Al atoms into the $\varepsilon$-Ge layer. As the abrupt $\varepsilon$-Ge/In$_{0.25}$Al$_{0.75}$As(001) exhibits a type-I band alignment, these calculations show that atomic-scale abruptness of this interface is required to achieve significant optical recombination in the $\varepsilon$-Ge layer, which hinders the use of this particular interface in optical devices.

\subsubsection{As diffusion into Ge} \label{sec-Asdiff}

Calculations of band offsets were also performed for the case of As-terminated In$_{x}$Al$_{1-x}$As, see Table~\ref{tab:AsGe_BO}.\footnote{For this case the As atoms of the MIML are moved into Ge in the manner described in sec.~\ref{sec-intdiff}, where the mixed layer stoichiometries are related by [As$_{0.5-−a}$Ge$_{0.5+a}$]$^\textrm{ML0}$ = [As$_{a}$Ge$_{1−-a}$]$^\textrm{ML1}$ for As atoms diffusing from ML0 to ML1, and [As$_{0.5-−b}$Ge$_{0.5+b}$]$^\textrm{ML1}$ = [As$_{b}$Ge$_{1−-b}$]$^\textrm{ML2}$ for As atoms diffusing from ML1 to ML2.} The results show that band alignments are quite sensitive to diffusion distance into Ge. For the case of As atoms residing in ML1 ([As$_{0.5}$Ge$_{0.5}$]$^\textrm{ML1}$), the VBO is reduced by 0.23 eV compared to the abrupt (ML0) case, while the CBO correspondingly increases by 0.24 eV. For $\varepsilon$-Ge/In$_{0.25}$Al$_{0.75}$As(001), the VBO (CBO) decreases (increases) by 0.14 eV (0.15 eV). When As atoms have diffused to ML2 ([As$_{0.5}$Ge$_{0.5}$]$^\textrm{ML2}$), the BOs continue to move in the same direction, showing again a linear change with respect to interface stoichiometry as in the case of group-III diffusion into Ge (Sec.~\ref{sec-catdiff}), but with a slope of the opposite sign. 

The VBO and CBO of Ge/AlAs for the [As$_{0.5}$Ge$_{0.5}$]$^\textrm{ML2}$ case (0.46 eV and 0.92 eV, respectively) compare very well with recent BO measurements of the Ge/AlAs interface.\cite{Hudait2014} The calculated VBO (CBO) of $\varepsilon$-Ge/In$_{0.25}$Al$_{0.75}$As(001) for [As$_{0.5}$Ge$_{0.5}$]$^\textrm{ML2}$ is 0.48 eV (1.15 eV). The latter BOs also compare well with unpublished experimental XPS measurements~\cite{Hudait-private} of the band alignment of $\varepsilon$-Ge/In$_{0.25}$Al$_{0.75}$As(001).\footnote{The manuscript reporting this joint experimental and theoretical effort is currently under preparation. We therefore omit the band offset figures for this case (refer to Fig.~\ref{fig:VBOs_CBOs_InxAl1-xAs_As-Ge-int} for the explicit models of [As$_{0.5}$Ge$_{0.5}$]$^\textrm{ML0}$ in Ge/AlAs(001) and $\varepsilon$-Ge/In$_{0.25}$Al$_{0.75}$As(001)).}

\subsubsection{Ge diffusion into In$_x$Al$_{1-x}$As} \label{sec-Gediff}

Due to the solid solubility of Ge in GaAs,\cite{Bosi2010} Ge diffusing through the interface towards the overlayer is a common observation in III-V/Ge(001) heterostructures (i.e. a III-Vs grown on Ge) such as GaAs/Ge(001). To a certain extent this diffusion and the overall interface quality can be controlled by growth conditions,\cite{Tanoto2006,Brammertz2008,Bosi2011,Sophia2015,Jia2016} as well as by thin interlayers of AlAs (or alloys thereof) between GaAs and Ge.\cite{Chia2008,Li2013,Qi2014,Chen2016} The latter technique can decrease interdiffusion due to the large Al-As bond energy and yield heterostructures with very sharp interfaces between the III-V region and Ge. However diffusion cannot be completely suppressed and Ge diffusion distances of a few nm to tens of nm into the AlAs region can be observed.\cite{Chia2008} Many factors influence the investigation of heterostructures involving III-Vs grown on Ge, such as the potential of CMOS compatible monolithic integration of optical devices\cite{Fitzgerald1992,Chilukuri2007} where graded GeSi alloys act as a buffer between the III-V overlayer and the Si substrate. Also, high-quality III-V/Ge interfaces with a type-I BO could offer advantages for photovoltaic technologies.\cite{Bosi2007,Guter2009,Qi2014} An understanding of the effects of Ge diffusion through the interface is imperative to the assessment of these potential applications.

In this section, the effects of Ge diffusion into AlAs and In$_{0.25}$Al$_{0.75}$As on the BOs are studied for short diffusion distances. For the interface with In$_{0.25}$Al$_{0.75}$As, Ge is tensile strained to the III-V lattice constant, which models the top In$_{0.25}$Al$_{0.75}$As/$\varepsilon$-Ge interface of a confined $\varepsilon$-Ge region between a III-V overlayer and a III-V substrate, as is considered for optoelectronic applications.\cite{Pavarelli13} Only the positions of Ge atoms which satisfy electron counting rules\cite{Peressi98,Martin80,Pashley89} for interfacial bonding are considered for this investigation, as this prevents the accumulation of an electric field across each slab which would result in unstable interface structures. For this reason, only the position of Ge atoms corresponding to the second monolayer away from the ML0 position and towards In$_{x}$Al$_{1-x}$As (ML-2) is considered. In terms of the stoichiometric expression for the mixed layer as defined at the beginning of Sec.~\ref{sec-intdiff}, this corresponds to only [Ge$_{b}$(III)$_{1−-b}$]$^\textrm{ML-2}$ ([Ge$_{b}$(As)$_{1−-b}$]$^\textrm{ML-2}$) being considered and compared to the abrupt [(III)$_{0.5}$Ge$_{0.5}$]$^\textrm{ML0}$ ([As$_{0.5}$Ge$_{0.5}$]$^\textrm{ML0}$) interfaces for group-III (group-V) terminated In$_{x}$Al$_{1-x}$As. 

\begin{figure}
{
\includegraphics[width=1.0\columnwidth]{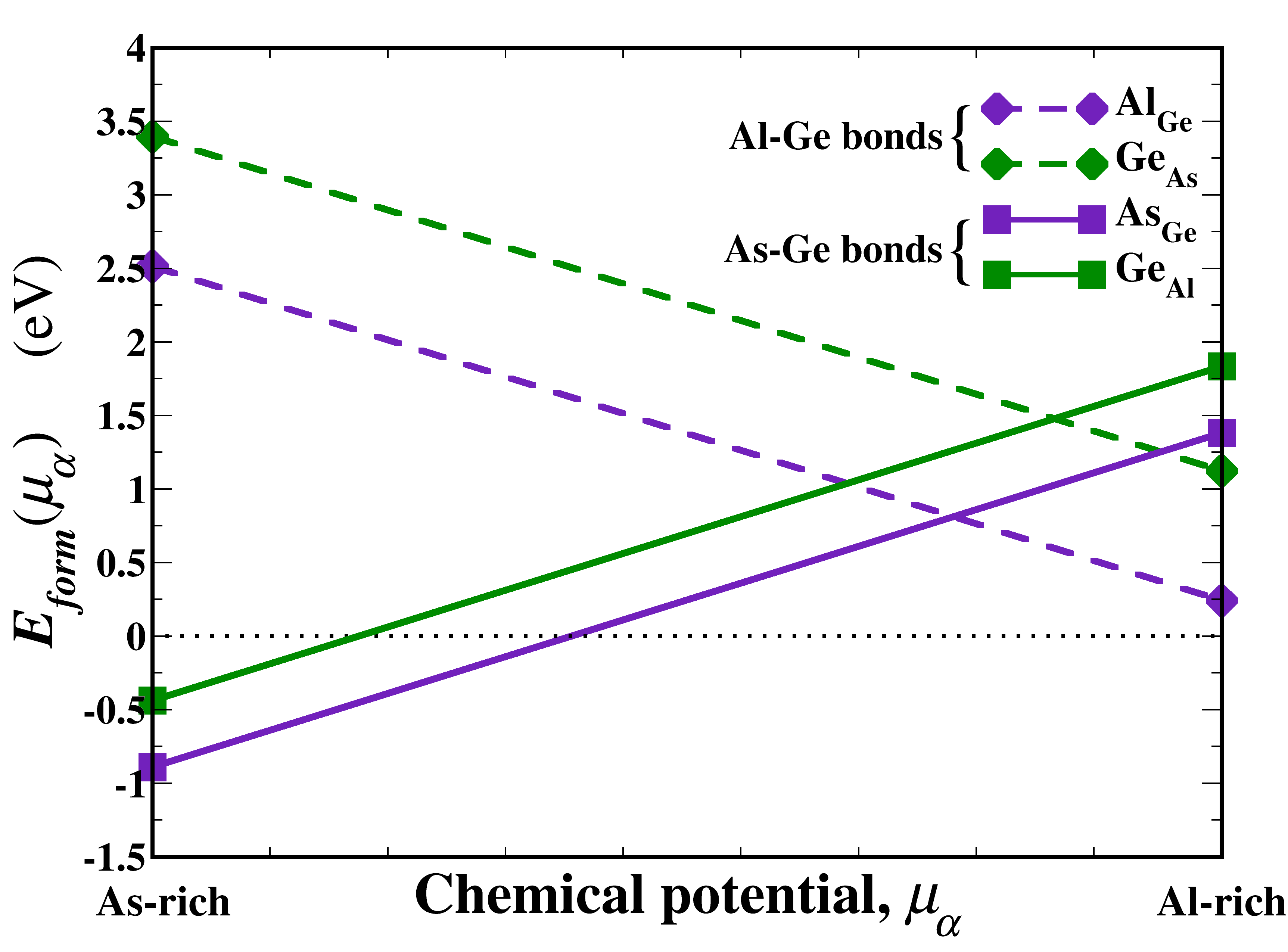}}
\caption{\label{fig:Eforms} Formation energies of substitutional impurities in Ge (purple lines) and impurities in AlAs (green lines), plotted as a function of chemical potential. Solid lines correspond to impurities which result in As bonding to Ge, while dashed lines correspond to impurities which result in Al bonding to Ge. In all cases, impurities which lead to As-Ge bonds have higher stability for the majority of the range of variation of the chemical potential, and particularly for As-rich conditions.}
\end{figure}

\begin{figure*}
{
\includegraphics[width=6.1225cm,height=4.75cm]{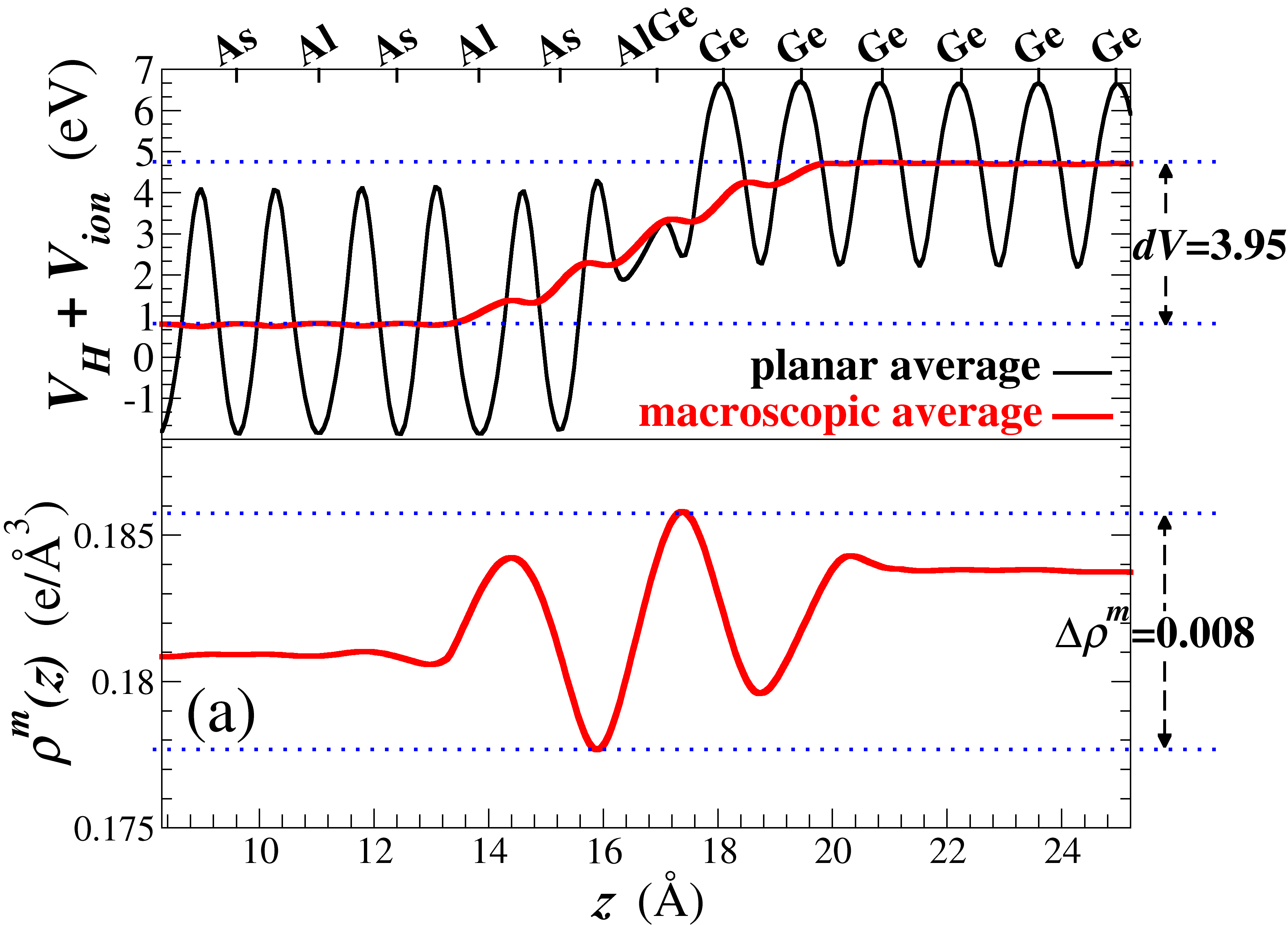} 
\includegraphics[width=5.8025cm,height=4.75cm]{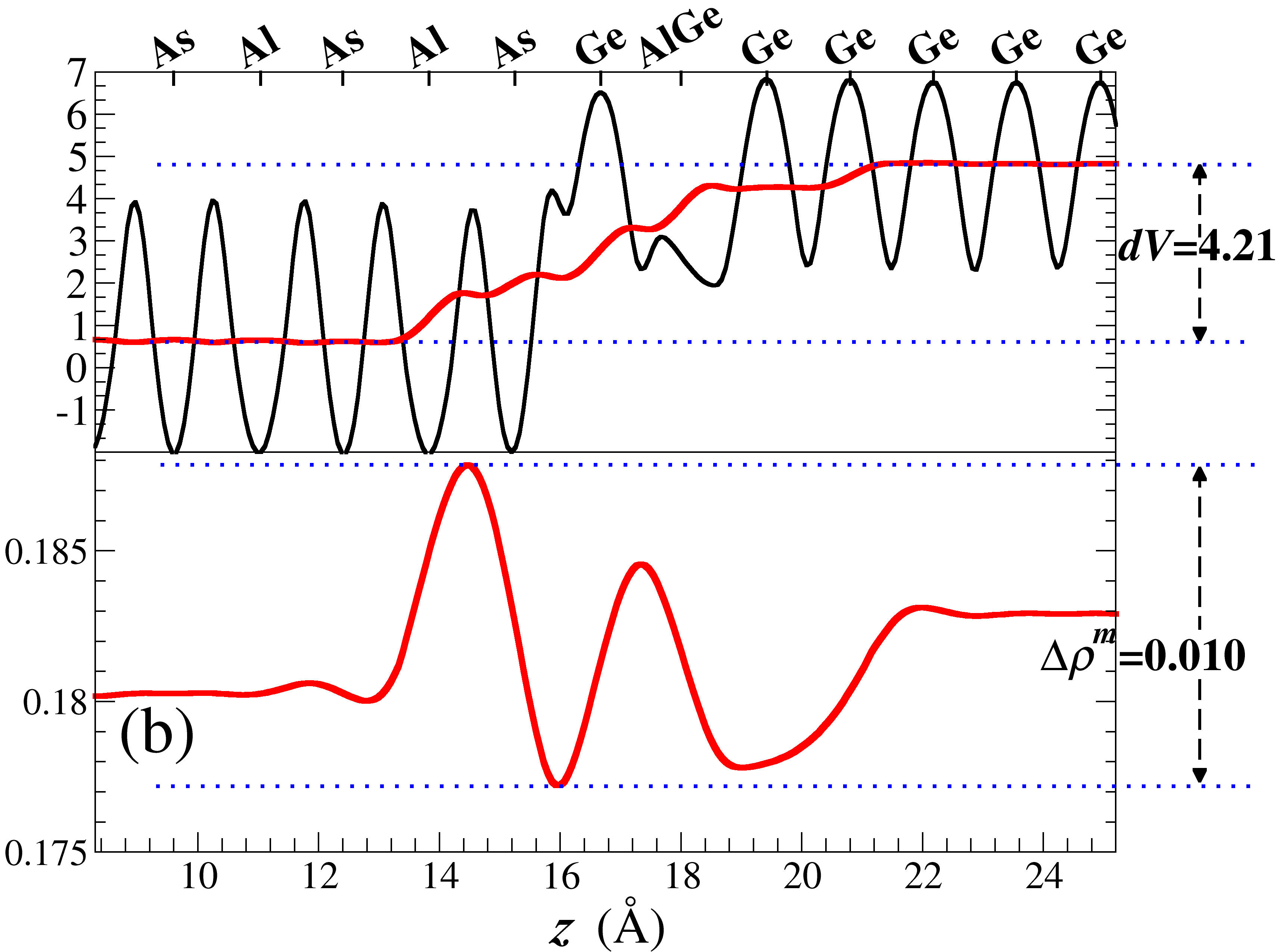}
\includegraphics[width=5.8025cm,height=4.75cm]{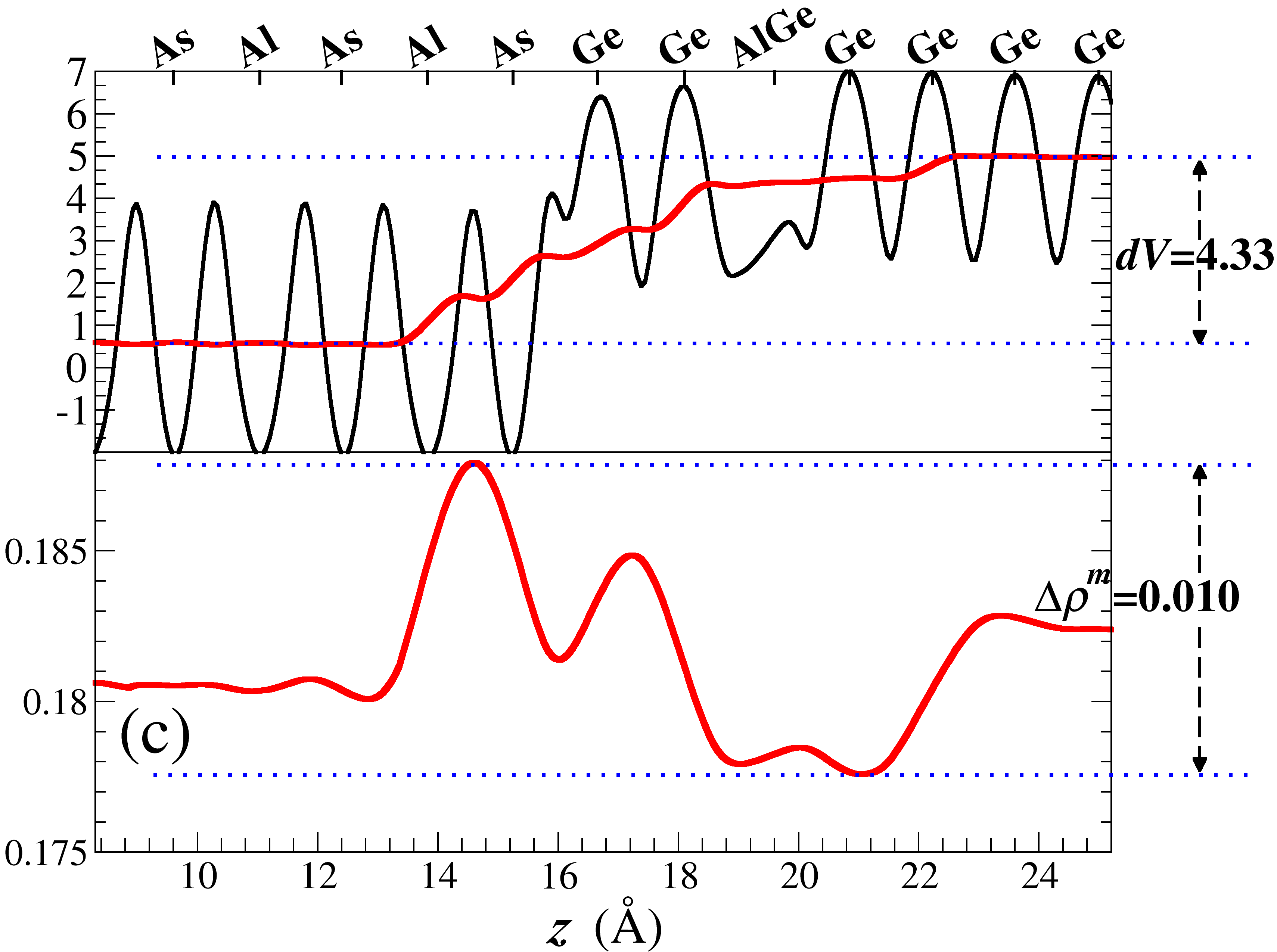}}

\caption{\label{fig:Velectro_rho_Aldiff} Al diffusing into Ge. Top graphs show the planar and macroscopic average of the electrostatic potential plotted as a function of $z$ (normal to interfacial plane). Bottom graphs show the macroscopic average of the electronic charge density plotted as a function of $z$. Left panels (a) correspond to the abrupt interface ([Al$_{0.5}$Ge$_{0.5}$]$^\textrm{ML0}$), middle panels (b) are for [Al$_{0.5}$Ge$_{0.5}$]$^\textrm{ML1}$, and right panels (c) correspond to Al residing two monolayers into Ge ([Al$_{0.5}$Ge$_{0.5}$]$^\textrm{ML2}$). All values of $dV$ are in eV. Charge density variation $\Delta$$\rho$$^{m}$ is in $e$/{{\AA}}$^3$. Note that the potentials and charge densities are plotted along the same horizontal scale.}
\end{figure*}

\begin{figure*}
{
\includegraphics[width=6.1225cm,height=4.75cm]{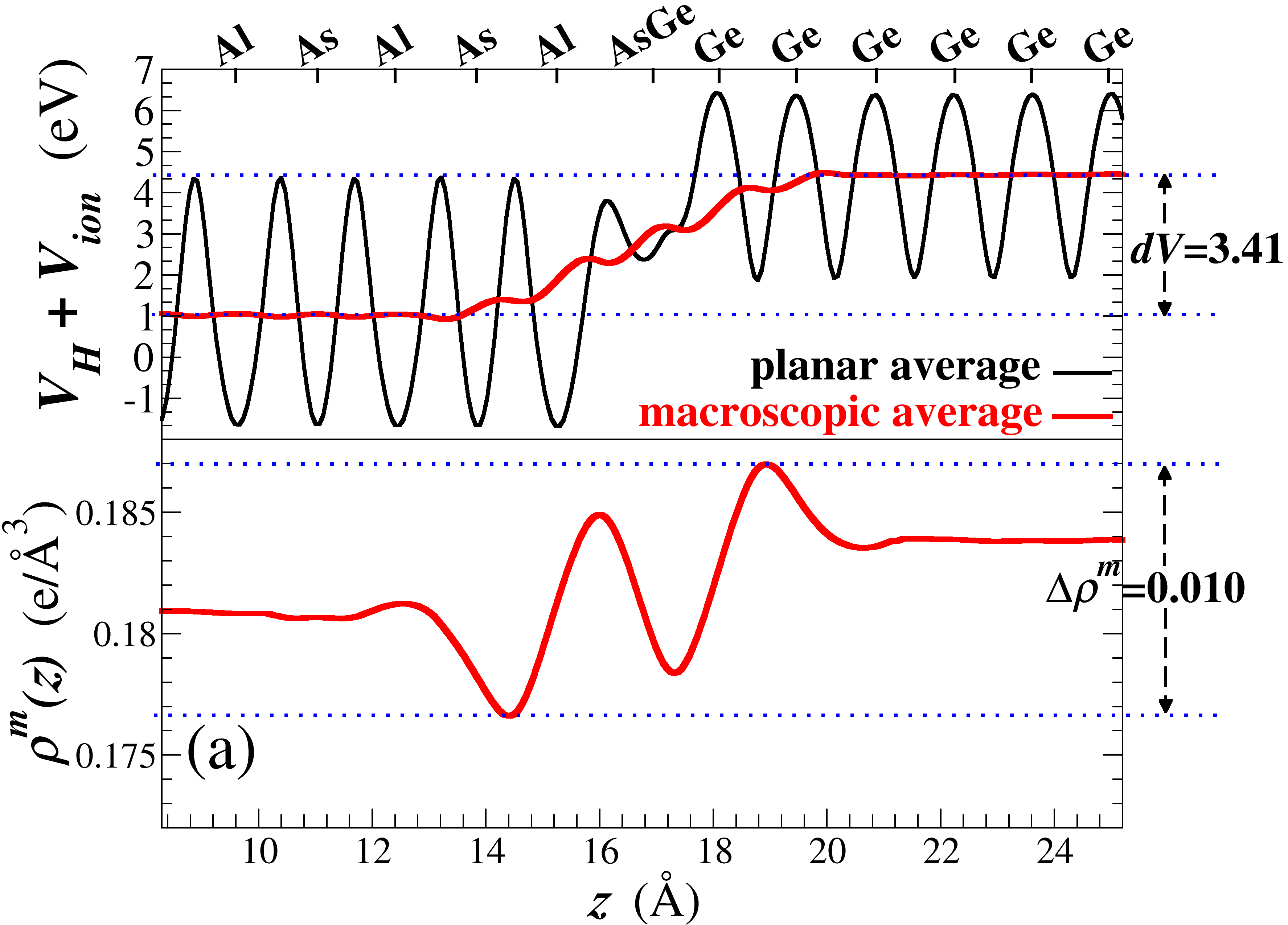} 
\includegraphics[width=5.8025cm,height=4.75cm]{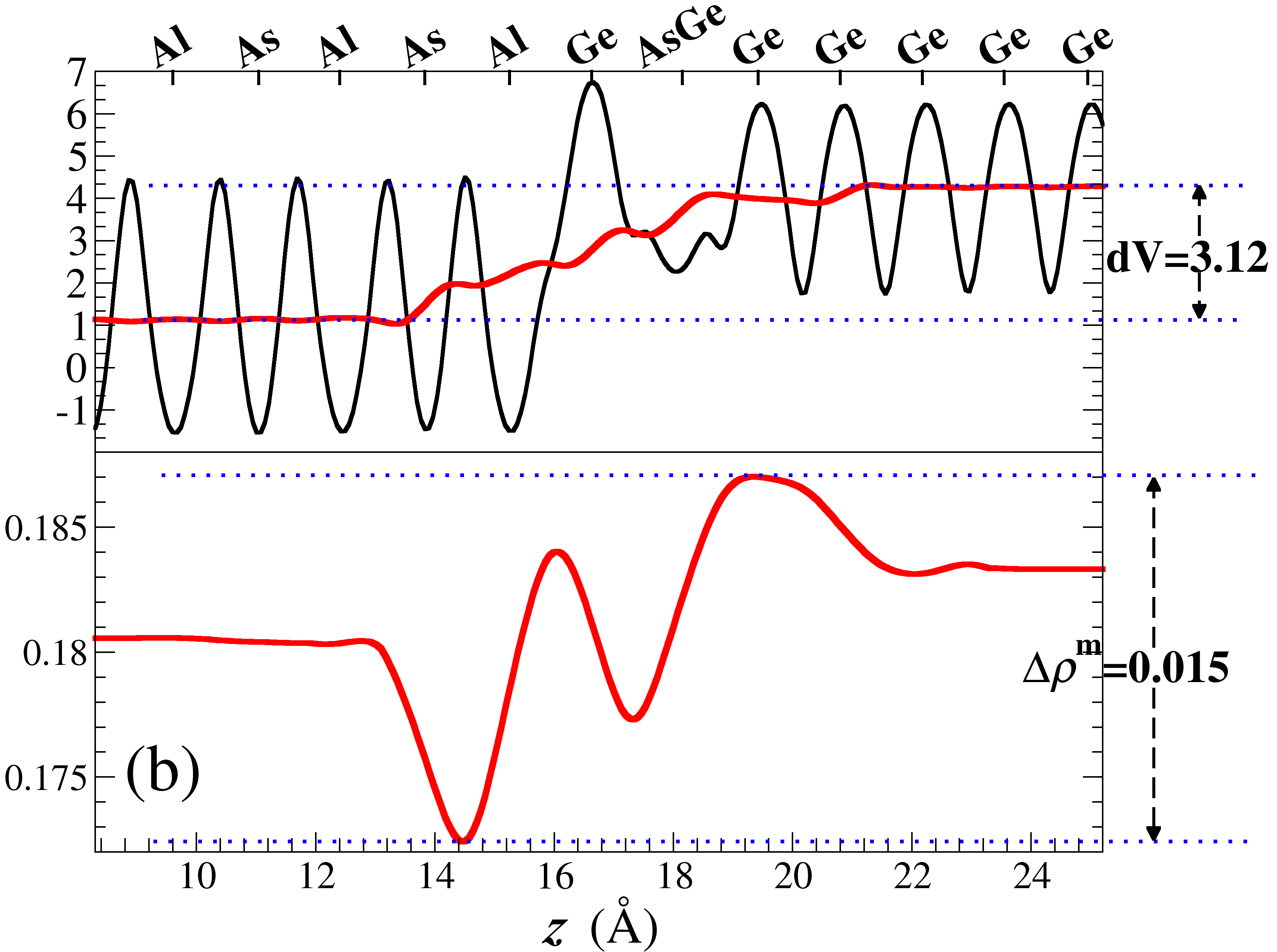}
\includegraphics[width=5.8025cm,height=4.75cm]{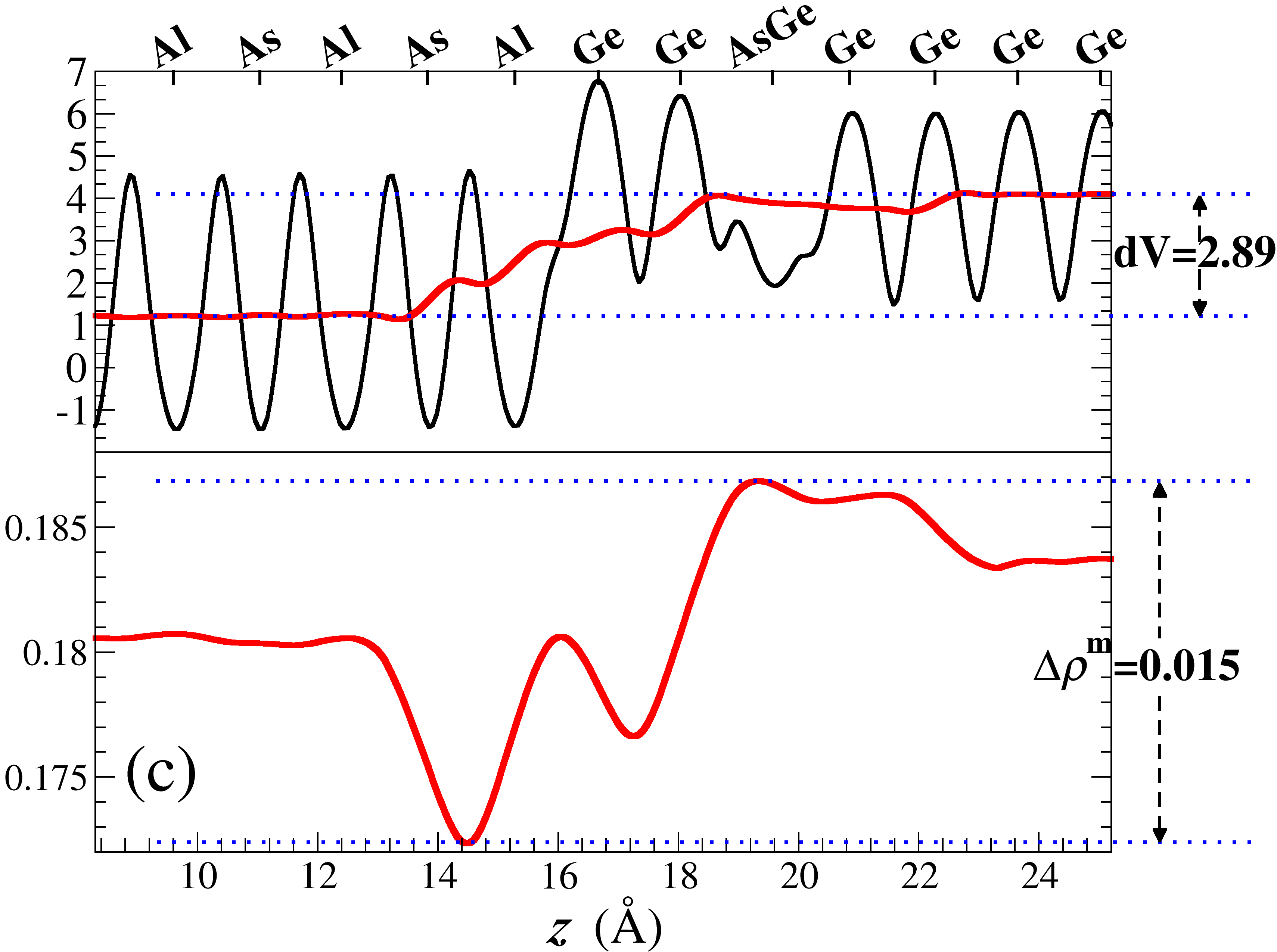}}

\caption{\label{fig:Velectro_rho_Asdiff} As diffusing into Ge. Top graphs show the planar and macroscopic average of the electrostatic potential plotted as a function of $z$ (normal to interfacial plane). Bottom graphs show the macroscopic average of the electronic charge density plotted as a function of $z$. Left panels (a) correspond to the abrupt interface ([As$_{0.5}$Ge$_{0.5}$]$^\textrm{ML0}$), middle panels (b) are for [As$_{0.5}$Ge$_{0.5}$]$^\textrm{ML1}$, and right panels (c) correspond to As residing two monolayers into Ge ([As$_{0.5}$Ge$_{0.5}$]$^\textrm{ML2}$). All values of $dV$ are in eV. Charge density variation $\Delta$$\rho$$^{m}$ is in $e$/{{\AA}}$^3$. Note that the potentials and charge densities are plotted along the same horizontal scale.}
\end{figure*}

\begin{figure*}
{
\includegraphics[width=8.9cm,height=6.6cm]{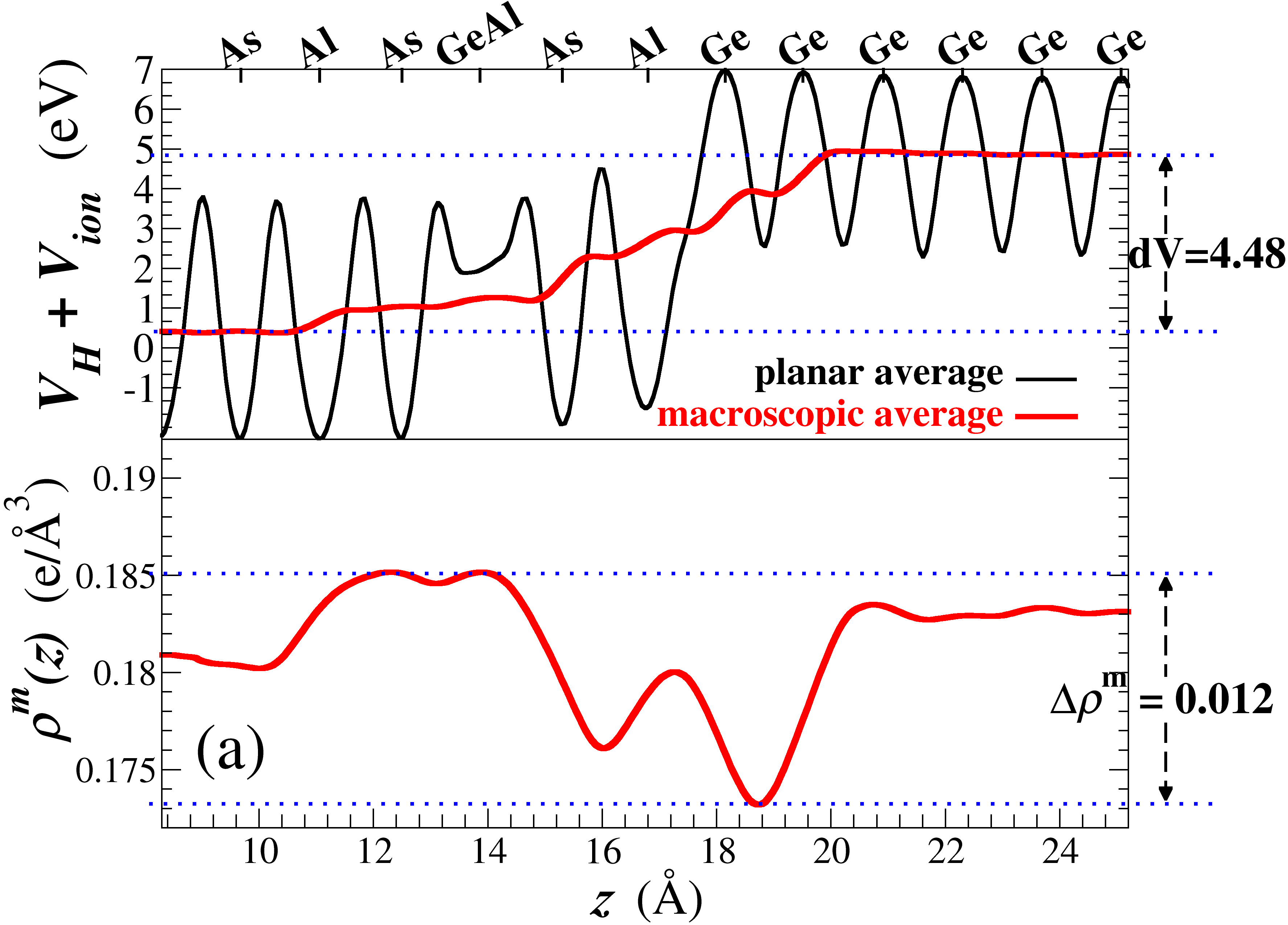} 
\includegraphics[width=8.4cm,height=6.6cm]{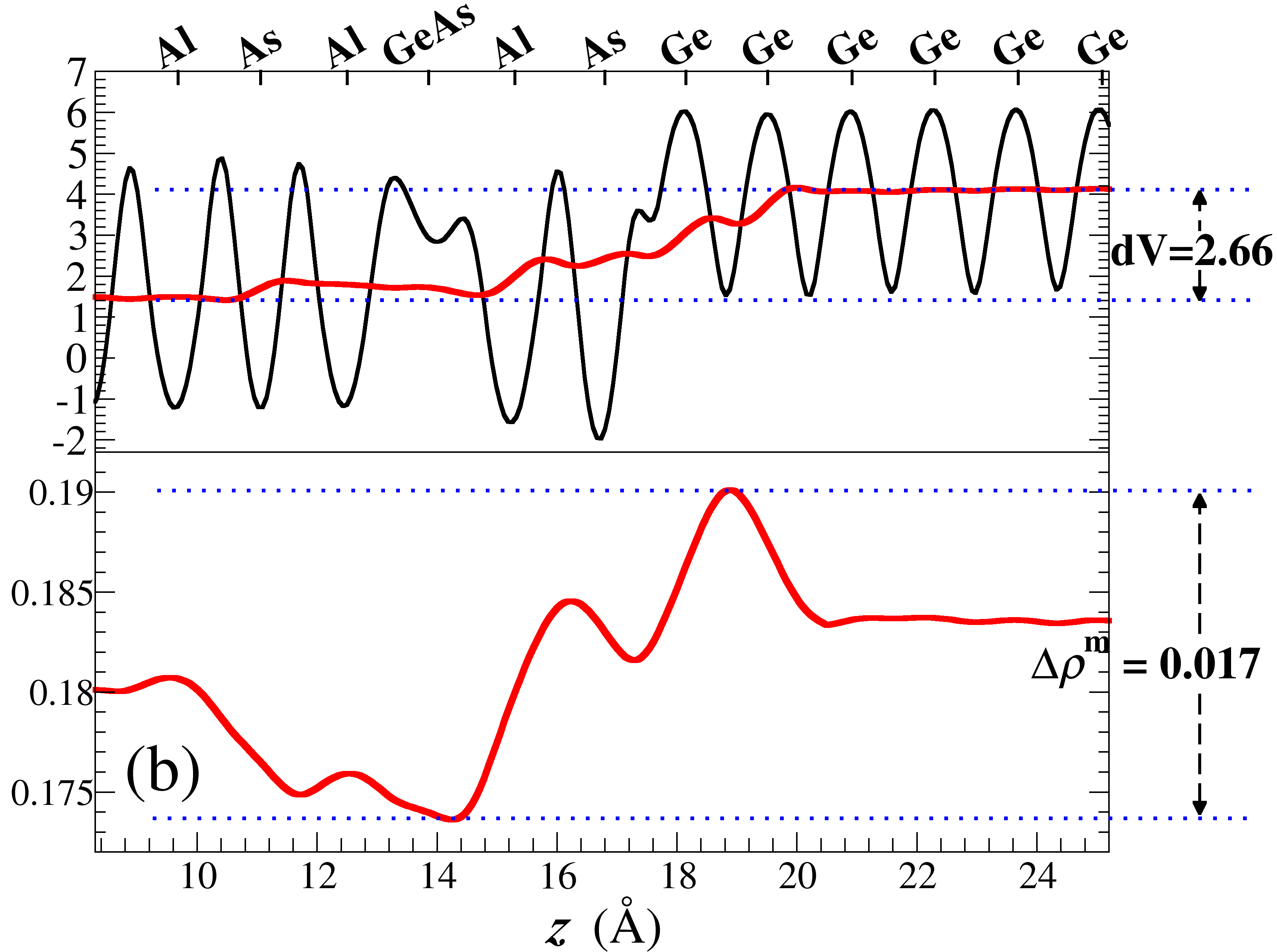}}

\caption{\label{fig:Velectro_rho_Gediff} Ge diffusing into AlAs. Top graphs show the planar and macroscopic average of the electrostatic potential plotted as a function of $z$ (normal to interfacial plane). Bottom graphs show the macroscopic average of the electronic charge density plotted as a function of $z$. Left panels (a) correspond to the Ge atoms residing two monolayers into Al-terminated AlAs ([Ge$_{0.5}$Al$_{0.5}$]$^\textrm{ML-2}$), and right panels (b) correspond to Ge residing two monolayers into As-terminated AlAs ([Ge$_{0.5}$As$_{0.5}$]$^\textrm{ML-2}$). All values of $dV$ are in eV. Charge density variation $\Delta$$\rho$$^{m}$ is in $e$/{{\AA}}$^3$. Compare $\Delta$$\rho$$^{m}$ of the left panel to that of [Al$_{0.5}$Ge$_{0.5}$]$^\textrm{ML2}$ in Fig.~\ref{fig:Velectro_rho_Aldiff}. Compare $\Delta$$\rho$$^{m}$ of the right panel to that of [As$_{0.5}$Ge$_{0.5}$]$^\textrm{ML2}$ in Fig.~\ref{fig:Velectro_rho_Asdiff}. Note that the potentials and charge densities are plotted along the same horizontal scale.}
\end{figure*}

In contrast to the results of Secs.~\ref{sec-catdiff} and ~\ref{sec-Asdiff} involving In$_{x}$Al$_{1-x}$As atoms diffusing into Ge, the band alignments in this section vary by a larger amount for a given diffusion distance. For group-III terminated In$_{x}$Al$_{1-x}$As (see Fig.~\ref{fig:Ge-diffusion_GecatML_Ge-InxAl1-xAs}), an increase of 0.55 eV (0.56 eV) in the VBO is calculated for Ge diffusing to the ML-2 position in AlAs (In$_{0.25}$Al$_{0.75}$As). For AlAs, this almost results in a broken gap band alignment, as the CBE of AlAs is only 0.09 eV above the VBE of Ge. The CBE of In$_{0.25}$Al$_{0.75}$As(001) is also close to the VBE of $\varepsilon$-Ge, but with a larger separation (0.12 eV) compared to Ge-AlAs. Assuming continued linearity of the BOs as a function of diffusion distance, broken gap alignments are expected for these interfaces with a further diffusion of Ge into In$_{x}$Al$_{1-x}$As. 

This has important consequences for devices involving AlAs (with possibly small proportions of In) grown on Ge. At variance, movement of the VBO in the opposite direction is calculated for Ge atoms diffusing into AlAs (In$_{x}$Al$_{1-x}$As) through an As terminated interface (see Fig.~\ref{fig:Ge-diffusion_GeAsML_Ge-InxAl1-xAs}). In this case the VBO \textit{decreases} by 0.75 eV (0.64 eV), which is qualitatively similar but significantly higher than the maximum VBO variation of 0.51 eV (0.38 eV) for As atoms diffusing through the Ge/AlAs(001) ($\varepsilon$-Ge/In$_{0.25}$Al$_{0.75}$As(001)) interface.  

An apparent bowing of the VBO as a function of the mixed layer stoichiometry $b$ can be observed in panels (c) and (d) of Figs.~\ref{fig:Ge-diffusion_GecatML_Ge-InxAl1-xAs} and~\ref{fig:Ge-diffusion_GeAsML_Ge-InxAl1-xAs}, whereas this bowing effect is largely suppressed for the case of group-III atoms diffusing into Ge (Fig.~\ref{fig:cat-diffusion_Ge-InxAl1-xAs} (c) and (d)). The bowing effect in Figs.~\ref{fig:Ge-diffusion_GecatML_Ge-InxAl1-xAs} and~\ref{fig:Ge-diffusion_GeAsML_Ge-InxAl1-xAs} is likely an artifact of the VCA model of the interfacial region; the VCA cannot correctly capture local structural properties\cite{Jaros85} which translates to errors in bond lengths and interlayer distances. These errors are exacerbated when the mixed monolayer, represented by the VCA, bonds to the neighboring ionic layers within the III-V crystal. The difference in bond lengths between III-V bonds and Ge-III/V bonds makes an important contribution to the local potential within the III-V slab and this contribution is missed in the VCA representation of the mixed layers for Ge diffusing into AlAs or In$_{0.25}$Al$_{0.75}$As. This causes larger errors for the intermediate values of $b$ which involve two types of VCA `atoms' in the supercell, thus producing the bowing effect. For group-III atoms diffusing into Ge (see Sec.~\ref{sec-catdiff}), the VCA sites are now bonding to covalent rather than to ionic layers and the structural errors of the VCA are not so apparent. 

The origin of the apparent bowing effect is further investigated by comparing the VCA results to explicit models of the mixed monolayer which are statistically representative of a 2 dimensional 50/50 random alloy (see panel (c) of Figs.~\ref{fig:cat-diffusion_Ge-InxAl1-xAs}, ~\ref{fig:Ge-diffusion_GecatML_Ge-InxAl1-xAs}, and ~\ref{fig:Ge-diffusion_GeAsML_Ge-InxAl1-xAs}). For the cases of Ge diffusing into AlAs, the VBOs obtained by the SQS monolayer are much closer to those obtained by the ordered model, compared to those obtained by the VCA. This lends further credence to the possibility that the apparent bowing effect seen in the VCA VBOs is due to structural errors associated with the VCA, as opposed to a realistic effect.
% since a poorer agreement is seen between VCA VBOs and VBOs from explicit models for interfaces which involve Ge diffusion into AlAs. 
This is also the case for Al diffusing one monolayer into Ge (see Fig.~\ref{fig:Ge-diffusion_GecatML_Ge-InxAl1-xAs}), although for two monolayers of Al diffusion the VBO from the ordered model resides midway between the VCA and SQS result. In addition, the reduced bowing in this case (Fig.~\ref{fig:Ge-diffusion_GecatML_Ge-InxAl1-xAs}) indicates that the VCA is a better approximation for modeling mixed layers in materials with purely covalent bonding, compared to materials with some degree of ionic bonding. 

%%\textit{...discussion of apparent bowing of valence band offsets with respect to interface stoichiometry...}

\subsubsection{Stability of diffused impurities} \label{sec-Eform}

The formation energies\cite{VandewalNeugebauer04,Freysholdt14,ZhangNorthrup91} of substitutional impurities were calculated in bulk cells in order to establish the relative stability (under conditions of thermodynamic equilibrium) of diffused impurities present in either Ge or AlAs after growth on top of either an AlAs substrate or a Ge substrate, respectively. Thus, the formation energetics establish, as a function of growth conditions, which diffused impurities are more likely to be present in either material after growth. The formation energies E$_{form}$($\mu$$_\alpha$) are plotted as a function of $\mu$$_\alpha$ for the each diffused impurity in Fig~\ref{fig:Eforms}. For the majority of the range of $\mu$$_\alpha$, the substitutional impurities which result in bonds between As and Ge have consistently lower formation energies than impurities corresponding to Al bonding to Ge. 

In particular, these As-Ge bonding impurities have very low formation energies (note that negative formation energies correspond to impurities which form spontaneously under thermodynamic equilibrium, hence a kinetic process would be required to prevent their formation) under As-rich conditions. The latter has been used to realize high quality, abrupt interfaces,\cite{Clavel2015,Nguyen2015} hence we expect As-rich growth conditions to be favored for applications of these heterostructures. We also mention that our calculated heat of formation of AlAs is $-$2.27 eV, which overestimates the magnitude compared to the experimental value of $-$1.53 eV\cite{Berger96semicon}. While this affects the formation energetics, our purpose is simply to establish an approximate, qualitative picture of the relative stabilities, and the error (0.74 eV) does not reverse the relative stability of As-Ge bonding impurities and Al-Ge bonding impurities under As-rich growth conditions.

\subsection{Analysis --- Relation between band offsets and interface configurations} \label{sec-analysis}
\subsubsection{Electrostatic potential, charge density, and interface diffusion} \label{sec-V_rho_diff}

The changes in band offsets presented in Sec.~\ref{sec-intdiff} arise purely from changes in the $dV$ term in Eqs.~(\ref{eq.1}) and (\ref{eq.2}). This is equivalent to stating that the changes in band offsets arise purely from interfacial effects, specifically from changes in the interface dipole derived from the macroscopic average of the atomic potentials near the interface, $V^{m}$$(z)$ (where the growth orientation is aligned to the $z$ direction). There is no contribution from bulk properties in the band alignment variations observed when comparing different interface structures (for a given group-III stoichiometry of In$_{x}$Al$_{1-x}$As and strain state of Ge).  

In general, atomic mixing affects the electrostatic potential line up by changing the charge density profile across the interface $\rho(z)$.\cite{Harrison78,Brillson16} In fact, it can be shown that the electrostatic potential (Hartree potential $V_{H}$ + bare ionic $V_{ion}$) step across the interface is given by
%the macroscopic average of the dipole moment generated by $\rho(z)$, 
\begin{equation}
  4\pi\,e^2\int\,z\rho^{m}(z)\text{d}z,
\end{equation}
and it is equivalent to the interface dipole ($e$ is the electronic charge, $\rho^{m}(z)$ is the macroscopic average of $\rho(z)$).\cite{Peressi98} Then, modifications to the local charge density arising from changes to the bonding configuration near the interface can provide either an enhancement or reduction of the interface dipole,\cite{Peressi98,McKinley92} depending on the polarity of bonds to the diffusing species, and their diffusion depth. For this reason, the VBO and CBO variations can be explained by electrostatic considerations involving the effect of positions of IV-III and IV-V bonds on the local potential.

As a result of the valence charge carried by the diffusing atoms, we expect Ge-III bonds to contribute positively to $dV$ as a function of diffusion distance of III atoms into Ge, and Ge-V bonds to contribute negatively to $dV$ as a function of diffusion distance. This is indeed consistent with what we observe in Figs.~\ref{fig:Velectro_rho_Aldiff} and~\ref{fig:Velectro_rho_Asdiff}. The former shows the planar and macroscopic averages of $V_{H}$+$V_{ion}$, and $\rho^{m}(z)$ for the explicit models of [Al$_{0.5}$Ge$_{0.5}$]$^\textrm{ML0}$, [Al$_{0.5}$Ge$_{0.5}$]$^\textrm{ML1}$, and [Al$_{0.5}$Ge$_{0.5}$]$^\textrm{ML2}$ for the Ge-AlAs interface; the latter shows the same quantities for the [As$_{0.5}$Ge$_{0.5}$]$^\textrm{ML0}$, [As$_{0.5}$Ge$_{0.5}$]$^\textrm{ML1}$, and [As$_{0.5}$Ge$_{0.5}$]$^\textrm{ML2}$ interface configurations of Ge/AlAs(001). It can be seen that as Ge-III (Ge-V) bonds move away from the abrupt interfacial layer and into Ge the step in the electrostatic potential increases (decreases), while the region over which charge transfer occurs widens. A similar conclusion is reached by plotting $V$$^{m}$$(z)$ and $\rho^{m}$$(z)$ for the $\varepsilon$-Ge/In$_{0.25}$Al$_{0.75}$As(001) interfaces, which are not shown for brevity.

For Ge diffusing into In$_{x}$Al$_{1-x}$As, (see Fig.~\ref{fig:Velectro_rho_Gediff}) a larger change in $dV$ relative to the abrupt interfaces is observed compared to the change in $dV$ for Al and As atoms diffusing the same distance into Ge. This is not unexpected, given the relation between $dV$ and $\rho^{m}$$(z)$\cite{Peressi98}; the variations in density across the interface $\Delta$$\rho$$^{m}$ have a slightly larger amplitude for Ge diffusion into Al- or As-terminated AlAs compared to [Al$_{0.5}$Ge$_{0.5}$]$^\textrm{ML2}$ or [As$_{0.5}$Ge$_{0.5}$]$^\textrm{ML2}$  (compare to corresponding $\Delta$$\rho$$^{m}$ values in Figs.~\ref{fig:Velectro_rho_Asdiff} and~\ref{fig:Velectro_rho_Aldiff}), and this translates to a larger effect on the interface dipole for Ge diffusion. 

\subsubsection{Linear response theory applied to interface diffusion} \label{LRT}
Here we derive a simple model to describe the relationship between interface dipole and mixed layer stoichiometry across a heterovalent interface. We follow the linear response theory approach\cite{Peressi98} put forward by Peressi {\it et al.} which is based on the model for polar interfaces proposed by Harrison $et$ $al$\cite{Harrison78}. Within this approach, the interface is treated as a perturbation of a periodic reference crystal and the potential lineup consists of an isovalent (i.e. interface independent) and heterovalent terms $dV$ = $dV_\text{iso}$ + $dV_\text{het}$. The $dV_\text{het}$ is then obtained via the Poisson equation from the additional nuclear charges (carried by the perturbation) at each site.
% and the interface stoichiometry by considering the group-IV/III-V interface being built up as a perturbation on a virtual zinc-blende crystal, with pseudovalence 3.5 and 4.5 for "cation" and "anion" sites respectively.

For Al diffusing into Ge, we consider bulk Ge as the reference crystal. Hence, the additional nuclear charge per site along each monolayer across the interface is (schematically) 
%the mixed layers containing diffused atoms, and the material's interface as a whole, as a perturbation on bulk Ge. 
\begin{equation}
\resizebox{\columnwidth}{!}{$\dots -\underbrace{\text{Ge}}_{0}-\underbrace{\text{Ge}}_{0}- \left\langle -\underbrace{\text{Ge}_{1-a'}\text{Al}_{a'}}_{(a')} \right\rangle - \left\langle \underbrace{\text{Ge}_{1-(a-a')}\text{Al}_{(a-a')}}_{(a-a')} \right\rangle - \left\langle \underbrace{\text{Ge}_{0.5+a}\text{Al}_{0.5-a}}_{0.5-a} \right\rangle - \underbrace{\text{As}}_{-1} - \underbrace{\text{Al}}_{+1}- \cdots$}
\end{equation}
where 0 $\leq a' \leq a \leq 0.5$ is the mixed layer stoichiometry. The accumulated charge, found by summing adjacent sites from left to right, is then used to find the net contribution of interfacial charge to $dV_\text{het}$, which for Al diffusing into Ge becomes
\begin{equation}
dV_\text{het}(\text{AlGe}) = \frac{\pi e^2}{2a_0\epsilon}(0.5 + a + a'),
\end{equation}
where $e$ is the electron charge, $a_0$ the lattice constant of a GeAlAs alloy\footnote{Since Ge and AlAs are lattice matched this will be the bulk lattice constant of either Ge or AlAs} and $\epsilon$ the dielectric constant of the same alloy obtained as an average of the Ge and the AlAs dielectric constant. This is similar for the case of As diffusing into Ge, but with the opposite sign,
\begin{equation}
dV_\text{het}(\text{AsGe}) = -\frac{\pi e^2}{2a_0\epsilon}(0.5 + a + a').
\end{equation}

For Ge diffusing into Al terminated AlAs, a similar line of reasoning results in the following interface contribution to the potential line up (again, $0 \leq b \leq 0.5$ is used here for the mixed stoichiometry instead of $a$)
\begin{equation}\label{eq:GeinAlAs}
dV_\text{het}(\text{GeAl}) = \frac{\pi e^2}{2a_0\epsilon}(0.5 + 2b),
\end{equation}

and for Ge diffusing into As terminated AlAs,

\begin{equation}\label{eq:GeinAsAl}
dV_\text{het}(\text{GeAs}) = -\frac{\pi e^2}{2a_0\epsilon}(0.5 + 2b).
\end{equation}

By considering $dV_\text{iso}$ as the average of the abrupt (ML0, $a$ = $a'$ = $b$ = 0) cases for Al and As terminated AlAs, the isovalent contribution to the VBO follows from Eq.~\ref{eq.1}. $dV_\text{het}$ for either Al(+) or As($-$) terminated ML0 cases is then obtained from the difference $dV$ - $dV_\text{iso}$ = $dV_\text{het}$, giving a value for the proportionality constant $\pi\,e^{2}/(2a_0\epsilon)=$~0.27~eV in Ge/AlAs and 0.245 eV in $\varepsilon$-Ge/In$_{0.25}$Al$_{0.75}$As. The contribution due to diffusing away from the ML0 plane is then simply an additional 0.27 (0.245) eV per monolayer of diffusion through the Ge/AlAs ($\varepsilon$-Ge/In$_{0.25}$Al$_{0.75}$As) interface, which results in a linear relation between the VBO and stoichiometry of mixed layers containing diffused impurities.\footnote{We have also derived $\pi\,e^{2}/(2a_0\epsilon)$ from first-principle by substituting in the expression the $a_0$ and $\epsilon$ values estimated from the LDA. In this case we obtained 0.29~eV and 0.27~eV respectively which would give comparable, though somewhat worse agreement with the values from the simulations.} 

Results show that this model agrees qualitatively with VBOs obtained from supercell calculations. We observe generally a better agreement between this model and calculations involving explicit interface configurations (non-VCA) than with VBOs obtained from the VCA representation of the interfaces (with the exception of Al diffusing 2 monolayers into Ge). The latter is particularly true for Ge atoms diffusing into (In)AlAs, which again shows the weakness of using VCA to represent mixed layers in a ionically bonded material.

\section{Conclusions} \label{conclusions}

First-principles calculations of valence and conduction band offsets have been performed using the DFT+$GW$ approach. The $GW$ correction was applied to obtain accurate bulk bandgaps, while DFT within the LDA formulation provided the interfacial profile of the self-consistent potential from which the interface dipole $dV$ can be derived. By varying the stoichiometry of monolayers near the interface using the VCA, the atomic diffusion away from the abrupt interfacial layer can be modeled. Within this approach, the $dV$ term can be changed depending on the interlayer stoichiometry with a sensitivity large enough to, in some cases, change the character of the band alignment. 

The results of this work are qualitatively consistent with the linear response theory developed for semiconductor interfaces,\cite{Resta89} where for heterovalent interfaces the change in interface dipole (and hence the change in band offset) should be linear in the stoichiometries of mixed layers.\cite{Peressi98,Bratina94} We attribute the deviations from linearity showing an apparent bowing effect, especially for the cases of Ge diffusing into the III-V slab to the structural errors associated with the VCA.

The VCA has also been used to model the group-III alloy of the III-V slab, thus introducing a further error in the calculations of the band offsets. This error has been investigated in Ref.~\onlinecite{Greene-Diniz16} for In$_{0.5}$Ga$_{0.5}$As.
%the movement of band edges due to alloying effects in the group-III sublattice of In$_{0.5}$Ga$_{0.5}$As\cite{Greene-Diniz16}.
%In this previous work, realistic models of the alloy were obtained from the SQS approach\cite{Zunger90,Wei90} and compared to the VCA.
By comparing with the most accurate SQS, it was found that most of the error in the VCA resides in the indirect L-point satellite valley band gap, while the minimum error of the VCA corresponds to the direct $\Gamma$-point band gap.
%within the VCA the direct $\Gamma$-$\Gamma$ band gap was overestimated by approximately 0.05 eV compared to the most accurate SQS model used, and the L-$\Gamma$  and X-$\Gamma$ conduction band gaps were underestimated by 0.22 and 0.10 eV, respectively. This shows that for the lowest lying conduction band satellite valleys, most of the error in the VCA resides in the indirect L-point satellite valley band gap, and the minimum error of the VCA corresponds to the direct $\Gamma$-point band gap.
As indicated by the early studies on SQS models, errors in band gaps obtained by averaging between the constituent binary materials generally follow the same trend for different III-V materials,\cite{Wei90} hence the trends in VCA-errors should be transferable to In$_{x}$Al$_{1-x}$As. For $x=25\%$---the highest In content alloy studied in the present work---the band gap is direct at the $\Gamma$-point, and we expect the least amount of error.
%(compared to the next lowest lying conduction band valleys) at the $\Gamma$-point band gap.
While this error is not negligible (likely $\lessapprox$ 0.1 eV), it is not large enough to change qualitative conclusions and trends of the present study. 

Future studies will involve a wider range of explicit models of disordered configurations for the interface, in which various SQS\cite{Zunger90,Wei90} representations of the interfacial layers will be compared against each other. This will shed more light on the band offset bowing effect and, by comparison, more accurately quantify the error in the band offsets when representing the mixed layer stoichiometries by the VCA. SQSs will also be used to model the group-III alloy of the III-V slab.

While the importance of the interface structure for heterovalent interfaces along with the associated departure from band offset transitivity seen for many isovalent interfaces is by now well-established, this work shows that variations in the interface stoichiometry can be enough to dramatically change the band alignment characteristics for the lattice (mis)matched ($\varepsilon$-)Ge/In$_{x}$Al$_{1-x}$As(001) interface. Combining this with the experimentally validated band offsets achievable from DFT+$GW$ for conduction and valence band offsets, this work shows that due to variations in the interface dipole, both type-I and type-II band offsets should be observable for this interface depending on the details of the interface structure. Calculations of the formation energetics of diffused substitutional impurities indicate consistently greater stability of impurities which involve As bonding to Ge, for both materials comprising the interface. For the commonly used experimental approach of growing Ge on As-rich (nominally As-terminated) III-As substrates, from which atomically sharp interfaces can be achieved, these results are consistent with the observation of type-I band offsets for ($\varepsilon$-)Ge/In$_{x}$Al$_{1-x}$As(001) for 0 $\leq$ $x$ $\leq$ 0.25. 

\vfill

\begin{acknowledgments}
  The authors thank J. C. Abreu for providing the SQS models. The authors are grateful for helpful discussions with J. C. Abreu, F. Murphy-Armando, T. J. Ochalski, D. Saladukha, M. B. Clavel, M. K. Hudait, and J. Kohanoff. 
  The authors acknowledge the use of computational facilities at the Atomistic Simulation Centre---Queen's University Belfast. This work is financially supported by the Department for the Employment of Northern Ireland and InvestNI through the US-Ireland R\&D partnership programme (USI-073).
\end{acknowledgments}

\bibliography{Ge-InAlAs}% Produces the bibliography via BibTeX.

\end{document}